\documentclass[prl,twocolumn,superscriptaddress,nofootinbib,amsmath,amssymb,longbibliography]{revtex4-1}
\usepackage{graphicx}
\usepackage{float}
\usepackage[usenames,dvipsnames]{xcolor}
\usepackage[colorlinks=true,citecolor=Blue,linkcolor=RubineRed,urlcolor=Blue]{hyperref}
\begin{document}

\title{
Using Active Matter to Introduce Spatial Heterogeneity to the Susceptible-Infected-Recovered Model of Epidemic Spreading 
}
\author{ P. Forg{\' a}cs}
\affiliation{Mathematics and Computer Science Department, Babe{\c s}-Bolyai University, Cluj-Napoca 400084, Romania}
\author{A. Lib{\' a}l}
\affiliation{Mathematics and Computer Science Department, Babe{\c s}-Bolyai University, Cluj-Napoca 400084, Romania}
\author{C. Reichhardt}
\affiliation{Theoretical Division and Center for Nonlinear Studies,
Los Alamos National Laboratory, Los Alamos, New Mexico 87545, USA}
\author{N. Hengartner}
\affiliation{Theoretical Division and Center for Nonlinear Studies,
Los Alamos National Laboratory, Los Alamos, New Mexico 87545, USA}
\author{C. J. O. Reichhardt$^*$}
\affiliation{Theoretical Division and Center for Nonlinear Studies,
Los Alamos National Laboratory, Los Alamos, New Mexico 87545, USA}

\date{\today}

\begin{abstract}
The widely used susceptible-infected-recovered (S-I-R) epidemic model assumes a uniform, well-mixed population, and incorporation of spatial heterogeneities remains a major challenge. Understanding failures of the mixing assumption is important for designing effective disease mitigation approaches. We combine a run-and-tumble self-propelled active matter system with an S-I-R model to capture the effects of spatial disorder. Working in the motility-induced phase separation regime both with and without quenched disorder, we find two epidemic regimes. For low transmissibility, quenched disorder lowers the frequency of epidemics and increases their average duration. For high transmissibility, the epidemic spreads as a front and the epidemic curves are less sensitive to quenched disorder; however, within this regime it is possible for quenched disorder to enhance the contagion by creating regions of higher particle densities. We discuss how this system could be realized using artificial swimmers with mobile optical traps operated on a feedback loop.
\end{abstract}
\maketitle

\section{Introduction}
Disease propagation through a heterogeneous environment has become a topic
of worldwide interest. Tremendous modeling resources have been applied in
efforts to control or at least predict the progress of the global pandemic. The
majority of these models have as their basis the conceptually simple yet
physically rich compartmentalized
susceptible-infected-removed (S-I-R) representation of temporal disease evolution
introduced nearly a century ago by Kermack and McKendrick \cite{Kermack27}.
Under the fundamental simplifying assumption of a mean-field, well-mixed
population, in the S-I-R model the population is divided into 
$S$ (susceptible), $I$
(infected), or $R$ (recovered) individuals, and the dynamic evolution of the 
epidemic is governed by the transition rates between these categories:
a removal rate $\mu$ for transitions I $\rightarrow$ R and an infection rate that relies
on on the law of mass action to model transitions from $S \rightarrow I$.
Individuals in a given bin are indistinguishable, and all spatial details
of the system are discarded \cite{PastorSatorras15}.
Despite their apparent simplicity, S-I-R models and their
many variants provide powerful tools for forecasting the general course of an
epidemic. Where these models falter is in predicting the specific course
of an actual real-world epidemic. This is generally attributed to the lack of
homogeneity in individual susceptibility, spatial contacts, and mixing
 behavior of individuals \cite{Hethcote00} 
 leading to stochastic effects that can not be averaged away.

Incorporation of heterogeneity has proven to be not at all straightforward, and
numerous approaches have been developed over the years. For example,
the population can be broken into subpopulations, each with different infection and recovery rates, or the 
population can be geographically subdivided into regions with 
 diffusive terms to link to the regional S-I-R dynamics 
\cite{Keeling99, Fenimore18}. Additional heterogeneities in the diffusion can be achieved
by incorporating patchiness 
into the diffusion \cite{Sun16,Kolton19}.    Much work has been done on connecting individuals
via finite-dimensional networks rather than through an infinite-dimensional mean
field \cite{PastorSatorras15}; however, the details of the network itself
make the problem even more complex since decisions must be made on
what
is the appropriate degree distribution for the network connectivity
as well as whether
the network should remain static or should be allowed to evolve either independently
or in response to the progress of the disease \cite{Volz07,Riley15}.
The impact of heterogeneity in transmission
and susceptibility is discussed in \cite{Miller07}.

The epidemic model with the ultimate heterogeneity
treats each individual as a separate,
mobile, interacting unit. Under Agent Based Modeling (ABM), 
also known as Individual Based Modeling \cite{Magal14}, heterogeneity can be included at
all levels ranging from varied individual susceptibility and recovery rates,
varied contacts between individuals, spatial clustering of individuals in cities
or at attractive sites such as bars, and both short and long range transport of
individuals such as by bus or airplane \cite{Eubank04,Germann06}.
The flexibility of these models is also their greatest weakness, since in addition
to the computational challenge of tracking
potentially  millions of individuals on a country-wide scale,
there can be a vast number of free parameters that must be painstakingly fitted
to real-world data that is not always available at the necessary resolution.

There have been surprisingly limited efforts to address a middle ground
of ABM in which many but not all of the details are abstracted away to produce
a model that captures spatial heterogeneity in a meaningful way without
being swallowed in a proliferation of parameters.  
This can, in principle,
be achieved either by developing more complex analytical models or 
simpler simulation-based models.  
One of the earliest
approaches for simplifying simulation-based models 
involved cellular automata, where the mobility of individual
agents could be varied up to a level consistent with the mean-field limit
\cite{Boccara92}.  Individuals obeying S-I-R interactions
have also been represented as moving particles that are driven and diffusing
\cite{Frasca06}, that never change direction \cite{Peruani08}, that
occasionally make long-range jumps \cite{Buscarino08},
that move at different velocities \cite{Rodriguez19}, or that are confined
to diffuse only within the region of their 'houses' \cite{Toledano21}.
To help mitigate the computational expense of such methods,
dynamic density functional theory
techniques can be applied \cite{teVrugt20}.

The significant progress made during recent
years in understanding what are
known as active matter models \cite{Marchetti13,Bechinger16}, where
individual particles are self-propelled and interact with each other on a spatial
landscape that may or may not include disorder, suggests the natural step of
pairing a model of S-I-R type with active particles.
The active particles can be of run-and-tumble type \cite{Peruani19} or
driven diffusive \cite{Paoluzzi20}. 
In a small system of low density, an active matter assembly was able to reproduce
the mean field behavior of S-I-R \cite{Norambuena20}. Generally, however, there
has been only limited work on coupling S-I-R modeling with active matter.
For example, Paoluzzi {\it et al.}
considered S-I-R type dynamics to examine information exchange in 
active clustering transitions 
\cite{Paoluzzi20} but not aspects of the epidemic spreading itself. 
Recently Zhao {\it et al.} studied
contagion dynamics in self-propelled flocking models and
found that ordered homogeneous states reduce disease spreading while
bands and clustering favor the spreading 
\cite{Zhao22}. 

There are
a number of advantages to working with an active matter system. The well-known
motility-induced phase separation (MIPS)
transition from a low density gas phase to a coexistence
between high and low density regions as a function of density and/or mobility of
the active particles
\cite{Fily12,Redner13,Palacci13,Cates15}
can provide a natural separation of the particles into
clustered communities connected by disordered transport pathways. Contacts
between particles can be viewed as an adaptive network that may be tuned to evolve
on the same or a different time scale as the progression of the disease. Spatial
heterogeneity emerges automatically in the MIPS regime, but can also be inserted
using walls, traps, or obstacles. 
Disease dynamics in such systems can be abstracted by tracking the evolution of the
number of $S(t)$, $I(t)$, and $R(t)$ over time, whose temporal behavior will capture
the impact of heterogeneities that are averaged out by the mean field approximations
of the standard S-I-R model.

In this work, we simulate a large assembly of active matter particles in the
MIPS regime where a giant cluster spontaneously emerges. We combine this
model with S-I-R interactions in which all particles are initially susceptible ($S$) but
can be infected with probability $\beta$ when they come into direct contact
with an infected ($I$) particle. Infected particles spontaneously transition to
the recovered ($R$) state at rate $\mu$, and no reinfection is allowed.
We study the evolution of epidemics
as the ratio of
$\beta/\mu$ is varied, and consider the impact
on the behavior of adding quenched disorder in the form of immobile obstacles.
Increasing the number of immobile obstacles
in an active matter system
will increase the number of clusters and decrease
their sizes.
By performing large numbers of realizations, we find that
inclusion of quenched disorder
increases the number of ``failed'' outbreaks for small $\beta/\mu$ and
increases the average duration of successful epidemics. When $\beta/\mu$
is sufficiently large, the system becomes insensitive to the presence of
quenched disorder and approaches the mean field limit,
and in this regime the epidemic propagates via spatially
well-defined fronts. We
also study the average
number of susceptible particles surrounding an infective as a function of time, and find
that this quantity is modestly altered by the addition of quenched disorder
in the mean field limit of high $\beta/\mu$ but becomes strongly
affected by quenched disorder
as $\beta/\mu$ is reduced.

Our results indicate that for low $\beta/\mu$, the
homogeneous
mixing
hypothesis breaks down,
that is, the infection process departs from mass action 
and the system becomes much more sensitive
to spatial quenched  disorder.
In the high $\beta/\mu$
regime, the mixing hypothesis is more applicable even though the epidemic
is spreading via spatially localized fronts.
This implies that localized epidemic mitigation efforts
will be more successful at low $\beta/\mu$ but would become ineffective
in higher $\beta/\mu$ regimes unless applied to the entire population.

Finally, we discuss how the system we consider
could be realized experimentally using
feedback control of light activated colloids,
where the active behavior of the colloids can be controlled on the
individual level.
Experiments on this type of system have already demonstrated
group formation, responsive states, and predator-prey
model realizations \cite{Lavergne19,Bauerle20,Chen22}.
There are also numerous possible ways to introduce
spatial heterogeneities in active matter systems
\cite{Reichhardt14,Lozano16,Pince16,Sandor17a,Bhattacharjee19a,Olsen21}.
Techniques of this type could be used to
mimic the S-I-R model with and without spatial disorder.
This could
permit the creation of table-top
epidemic spreading models with active matter.

\section{Results}
\begin{figure}
  \includegraphics[width=0.5\textwidth]{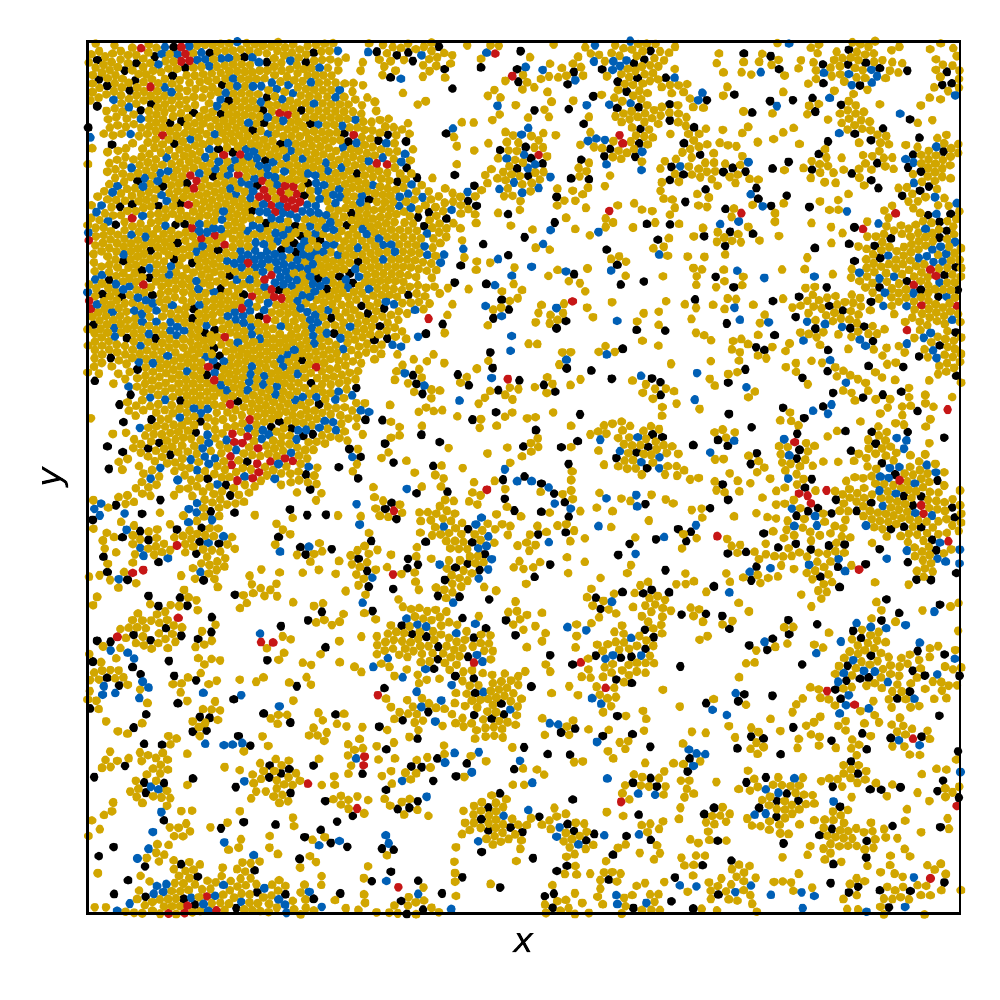}
  \caption{
{\bf Image of the sample containing run-and-tumble S-I-R particles in a motility-induced
  phase separated regime.}
The particles transition among
susceptible ($S$, yellow), infected ($I$, red), and recovered ($R$, blue) states.
Here $\beta/\mu=0.5$
and quenched disorder is present
in the form of $N_{\rm obs}=800$ immobile obstacles (black).
The quenched disorder causes the formation of numerous small clusters in addition
to the giant MIPS cluster.
  }
  \label{fig:1}
\end{figure}

\noindent{\textbf{\textsf{Modeling and characterization of the S-I-R dynamics}}}\\
We simulate $N=5000$ active particles in a two-dimensional system of
size $L \times L$ where $L=200$ and where there are periodic boundary
conditions in the $x$ and $y$ directions.
The motion of the particles is obtained by integrating the following equation:
\begin{equation} 
\alpha_d {\bf v}_{i}  =
{\bf F}^{dd}_{i} + {\bf F}^{m}_{i} + {\bf F}^{obs}_{i} \ .
\end{equation}
Here ${\bf v}_{i} = {d {\bf r}_{i}}/{dt}$ is the velocity and  
${\bf r}_{i}$ is the position of particle $i$, and the damping constant $\alpha_d = 1.0$.
The interaction between two particles is represented with a harmonic
repulsive potential
${\bf F}^{dd}_{i} = \sum_{i\neq j}^{N}k(2r_{a} - |{\bf r}_{ij}|)\Theta( |{\bf r}_{ij}| - 2r_{a}) {\hat {\bf r}_{ij}}$,
where $\Theta$ is the Heaviside step function,
${\bf r}_{ij} = {\bf r}_{i} - {\bf r}_{j}$, and
$\hat {\bf r}_{ij}  = {\bf r}_{ij}/|{\bf r}_{ij}|$. We set
the spring force to $k = 20$ and the particle radius to $r_{a} = 1.0$.
Each particle is subjected to a motor force
${\bf F}_i^m=F_{M}{\bf \hat{m}}_i$ of magnitude $F_M$ applied in
a randomly chosen direction ${\bf \hat{m}}$
during a run time of $\tau_{l}$ before instantaneously changing to a new
randomly chosen direction during the next run time, producing
run-and-tumble dynamics.
For each particle, 
we fix $\tau_{l}$ to a value selected randomly
from the interval $1.5\times 10^4$ to $3.0 \times 10^4$, and we set
$F_M=1.5$ for susceptible and recovered particles,
placing us in the MIPS regime in the absence
of quenched disorder \cite{Sandor17a}.
Infected particles have their motor force reduced to $F_M=1.0$.
For some simulations, we include quenched disorder in
the form of $N_{\rm obs}=800$ obstacles
that produce the force ${\bf F}^{obs}$. This is taken to be the
same as the particle-particle interaction force, with the only difference being
that the obstacles are immobile. An image of the system in the presence
of obstacles appears in Fig.~\ref{fig:1}.

Each active particle carries a label marking it as being in one of three states: $S$, $I$,
or $R$. If an $S$ particle comes into direct contact with an $I$ particle, for each time
step during which the contact persists there is a probability $\beta$ that
the $S$ particle will transition to an $I$ particle. If at a given time step an $S$ particle is
in contact with $n$ $I$ particles, the probability of infection is $1-(1-\beta)^n \approx n\beta$. Transitions
of $I$ particles to state $R$ occur with probability $\mu$ at each time step
regardless of the state of any particles that may be in contact with the $I$ particle.
Thus, the mean time spent in the infected state is
$1/\mu$ time steps.
The $R$ state is absorbing and $R$ particles experience no further state transitions.
In this S-I-R model, the infected $I$ particles are present only as a transient and the system will
eventually contain only $S$ and/or $R$ particles.
We note that the mean-field rates governing $S\rightarrow I$ and $I\rightarrow R$
transitions and determining the basic reproductive number $R_0$ in classic
S-I-R models do not map directly to the values of $\beta$ and $\mu$ that we
insert into our model as microscopic parameters. In ABMs, the
effective mean-field rates are emergent quantities instead of control
parameters.

To initialize the system, we place the particles randomly in the sample and set
them all to state $S$. We allow the system to evolve for
$5 \times 10^5$ simulation time steps
until a stable MIPS giant cluster has formed, and define this state to be the $t=0$
condition. We then randomly select 5 particles and change their state to $I$. These
particles serve as our index cases, and we choose 5 rather than 1 in order to
lower the probability of a failed outbreak. The system continues to evolve under
both the motion of the particles and the reactions between states $S$, $I$, and $R$ until
no $I$ particles remain.
We perform 1000 realizations for each parameter set. Since, as is shown in the
results,
the duration $t_d = \min\{t> 0: I(t)=0\}$ of an individual epidemic
can vary significantly from run to run, we report
time in terms of the scaled quantity $\tilde{t}=t/t_d$.
As a function of scaled time, we measure the epidemic
curves $s(\tilde{t})=S(\tilde{t})/N$, $i(\tilde{t})=I(\tilde{t})/N$, and
$r(\tilde{t})=R(\tilde{t})/N$.  
This rescaling enables us to visually compare features of the progression of the 
epidemic when we change the ratio $\beta/\mu$.
We also measure the
peak infective fraction $i_{\rm max}$ and the final susceptible
fraction $s_{\infty}$, both of which are commonly used indicators of the severity
of an epidemic. To obtain further information on the spatial evolution of the
system, we measure the average number of susceptible particles surrounding an
infective, $\eta(\tilde t)=I^{-1}(\tilde t)\sum_{i}^{I(\tilde t)}\sum_{j}^{S(\tilde t)} {\mathbb I}(|r_{ij}(\tilde t)| = 2r_a)$,
where ${\mathbb I}$ denotes the indicator function, and the sums over $i$ and $j$ range over the
infected and susceptible particles, respectively.
For a two-dimensional system of particles with identical radii $r_a$, $\eta$ cannot exceed
the maximum coordination number of $z=6$.  
If the infected individuals are well mixed within the population, the average number
of susceptible particles surrounding an infective
is $\eta(\tilde t) \propto S(\tilde t)$; more specifically, we would
expect $\eta(\tilde t) \propto z_c S(\tilde t)/N$, where $z_c$ is the average coordination
number of the particles.  Departure from
this behavior
is indicative of a failure of the homogeneous mixing assumption.

\begin{figure}
  \includegraphics[width=0.5\textwidth]{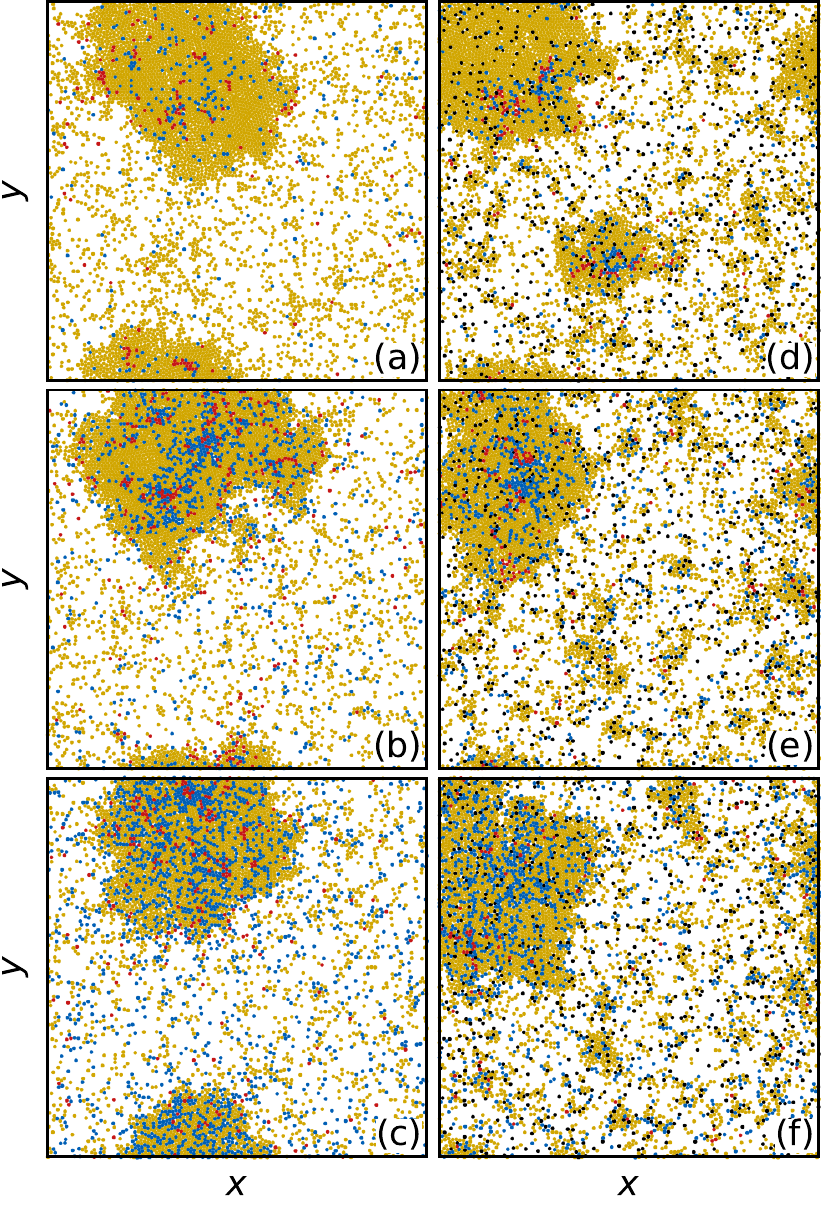}
  \caption{
{\bf Images of the low transmissibility regime with and without quenched disorder.}    
The evolution of the epidemic for the system in Fig.~\ref{fig:1} with
$\beta/\mu = 0.5$ at
time (a,d) $\tilde{t}=0.2$, (b,e) $\tilde{t}=0.3$, and (c,f) $\tilde{t}=0.4$.
The particles transition among
susceptible ($S$, yellow), infected ($I$, red), and recovered ($R$, blue) states.
(a,b,c) The obstacle-free system.
(d,e,f) The system containing obstacles, showing that fewer infected particles
are present at later times.
Movies of these two systems are available in the Supplemental Material.
  }
  \label{fig:2}
\end{figure}

\noindent{\textbf{\textsf{Low Transmissibility Regime}}}\\
In Fig.~\ref{fig:2}(a) we show a snapshot of
the system at $\beta/\mu=0.5$ in the low transmissibility regime
in the absence of quenched disorder.
The moving particles form a phase separated state of a high density
solid and a low density gas.
As indicated in the introduction, the relationship between $\beta/\mu$ and the
basic reproductive number is an emerging quantity.  Since within a cluster
the expected number of contacts is $z=6$, the expected number of secondary cases
from an index case within the cluster will be $\eta=3$,
showing that the epidemic will infect
a fraction of the cluster.  If the index case starts
in the gaseous phase, its expected 
number of contacts is likely $z<1$,
implying a reproductive number
less than one and giving limited cluster-to-cluster transmissions.

The same system in the presence of randomly placed obstacles
appears in Fig.~\ref{fig:2}(b),
where the giant dense cluster is now accompanied by numerous smaller persistent clusters
that have nucleated around some of the obstacle sites.  
Since particles within a cluster are
locked to one another, the homogeneous mixing assumption fails to hold for the epidemic dynamic within
each cluster.   Thus by controlling the number and size of the clusters, we explore a range
of departures from the mixing assumption,
from the most extreme situation when there is only
one large cluster, to greater mixing
as the number of clusters increases and their sizes decrease.

In Fig.~\ref{fig:2}(a,b,c) we illustrate
the evolution of the $S$, $I$, and $R$ particles
for the obstacle-free system at times of
$\tilde{t}=0.2$, 0.3, and 0.4,
while
in Fig.~\ref{fig:2}(d,e,f)
we show the evolution in the system containing $N_{\rm obs}=800$ obstacles.  
In both cases, when the giant cluster is
contacted by an infective, the disease spreads through the cluster, but since
the probability of transmission is low, not all of the $S$ particles surrounding
a given $I$ become infected, 
and as a result, a finite number of $S$ remain when the epidemic is complete.
See Ref.~\cite{Miller12} for a discussion on final epidemic size.
When we add quenched disorder to the system, shown as black circles in
Fig.~\ref{fig:2}(d-f),
a greater amount of localized clustering occurs
in addition to the giant cluster.
Since each cluster must be infected separately, this tends to
slow the spread of the infection and reduce
the peak infective fraction $i_{\rm max}$,
as shown in Fig.~\ref{fig:2}(e).

Although the dynamics of the spread of the infection is similar
with and without quenched disorder,
at $\tilde{t}=0.4$
the number of $I$ particles present
is much lower when obstacles have broken the system into smaller
clusters, indicating that the 
epidemic has
impacted fewer particles in the system with quenched disorder.

\begin{figure}
  \includegraphics[width=0.5\textwidth]{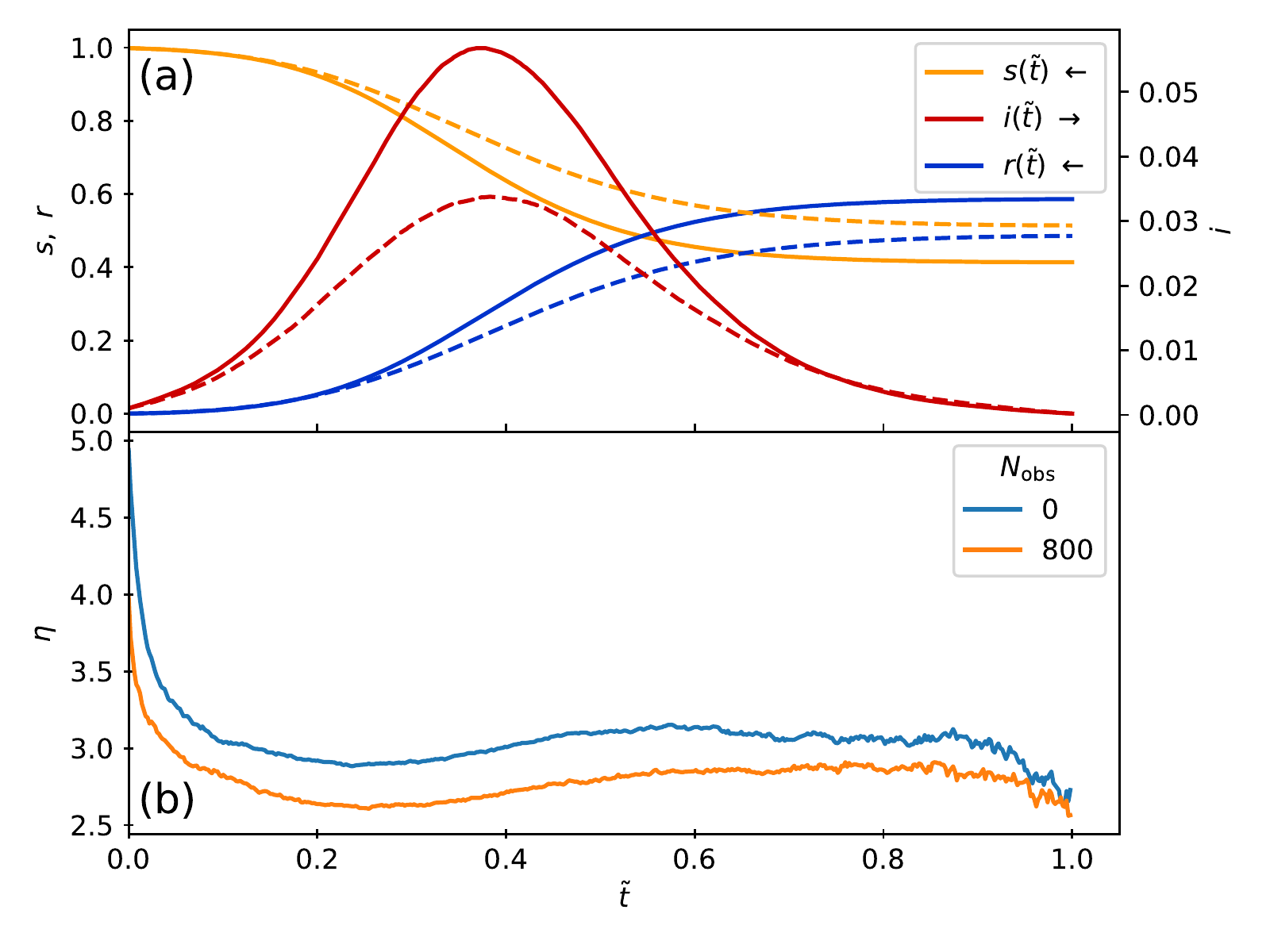}
  \caption{
{\bf Epidemic curves in the low transmissibility regime.}    
(a) Fractions of susceptible $s$ (yellow), infected $i$ (red), and recovered $r$ (blue)
particles versus rescaled time $\tilde{t}$ for the system in Fig.~\ref{fig:2} with
$\beta/\mu=0.5$. Solid lines are for samples without quenched disorder and dashed
lines are for samples containing obstacles.
At $\tilde{t} = 1.0$, the epidemic is over and $i=0$.
Introducing obstacles reduces the peak value $i_{\rm max}$ of the infective curve.
(b) The corresponding $\eta$, the average number of $S$ particles surrounding
an $I$ particle, versus $\tilde{t}$ in the sample without (blue) and with (orange)
obstacles.
Here, the inclusion of obstacles significantly reduces $\eta$ during the
entire epidemic.
  }
  \label{fig:3}
\end{figure}

In Fig.~\ref{fig:3}(a) we plot the epidemic curves showing the
fractions of susceptible $s$,
infected $i$, and recovered $r$ particles as a function of reduced
time $\tilde{t}$ for samples with and without quenched disorder.
We note that in the presence of obstacles, the duration of the epidemic
tends to be longer; however, by plotting the epidemic curves as a function of
reduced time it is easier to compare samples with and without quenched
disorder.
The curves have the shapes expected from the classic S-I-R model.
In the system without obstacles,
by the end of the epidemic there is still a fraction $s_{\infty}=0.41$ of the population
that never became infected,
while $r_{\infty}= 1 - s_\infty =0.59$ of the particles have recovered.
When obstacles are present, a larger fraction $s_{\infty}=0.51$
of particles have escaped infection.
The peak $i_{\rm max}$ in the infected fraction
is also considerably reduced in magnitude
when obstacles are present.
This indicates that the system is sensitive to the
presence of spatial heterogeneities introduced by the clustering 
arising from the presence of fixed
obstacles.
Within this regime, spatially localized mitigation protocols could be
effective, since local quenched disorder can
slow the overall mobility of the particles or reduce the effective
connectivity among the particles.
To further demonstrate the effect of adding obstacles,
in Fig.~\ref{fig:3}(b) we plot $\eta$, the average number of $S$ particles surrounding
an $I$ particle, versus $\tilde{t}$.
Here $\eta$ is always smaller in the sample containing obstacles.

\begin{figure}
  \includegraphics[width=0.5\textwidth]{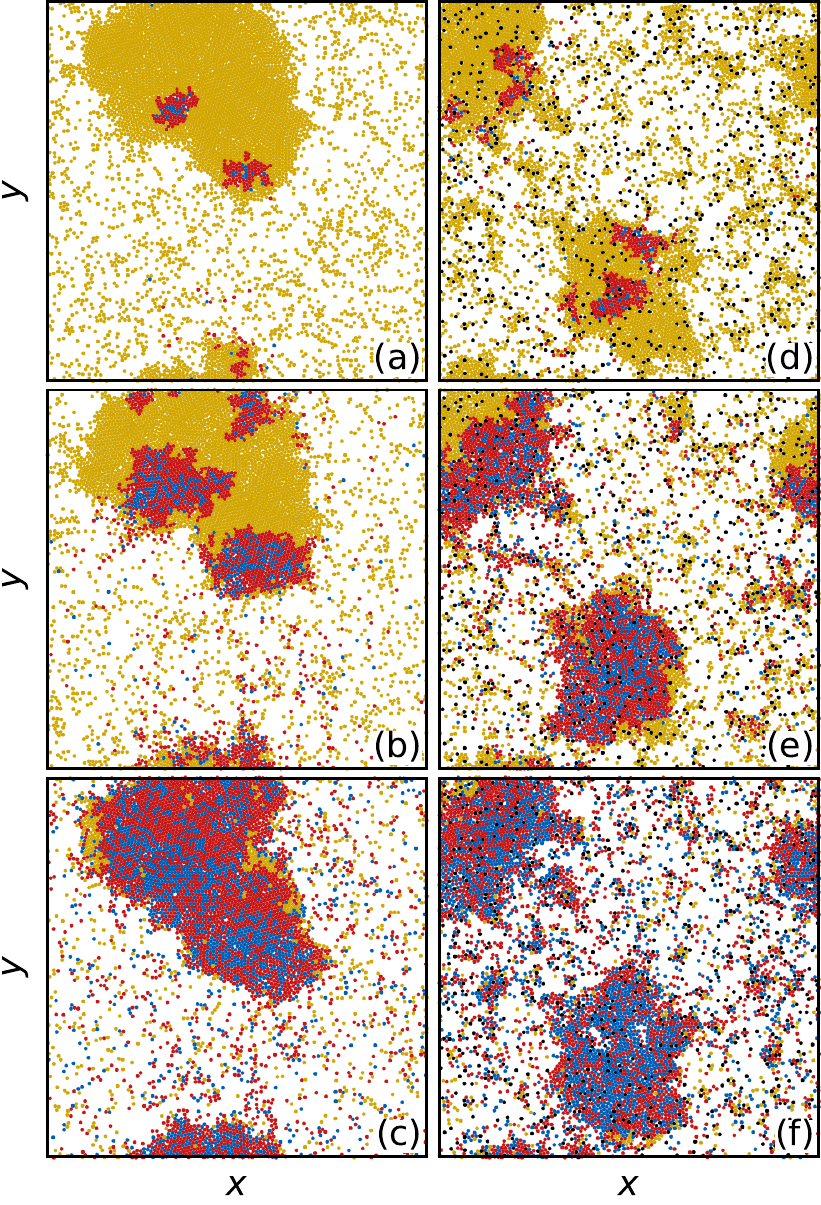}
  \caption{
{\bf Images of the high transmissibility regime with and without quenched disorder.}    
The evolution of the epidemic for systems with $\beta/\mu = 5.0$
at time (a,d) $\tilde{t}=0.1$, (b,e) $\tilde{t}=0.2$, and (c,f) $\tilde{t}=0.3$.
The particles transition among
susceptible ($S$, yellow), infected ($I$, red), and recovered ($R$, blue) states.
(a,b,c) The obstacle-free system. (d,e,f) The system containing obstacles.
For both cases, the epidemic spreads as a well-defined front through the
dense clusters.
Movies of these two systems are available in the Supplemental Material.
  }
  \label{fig:4}
\end{figure}

\noindent{\textbf{\textsf{High Transmissibility Regime}}}\\
We next consider the case of high transmissibility $\beta/\mu=5.0$.
In Fig.~\ref{fig:4}(a,b,c) we plot the spatial evolution of the susceptible,
infected and recovered particles
in the absence of obstacles. The infection spreads via
well defined fronts through the dense region.
In Fig.~\ref{fig:4}(d,e,f) we show the same evolution in the presence of
obstacles. There are now multiple dense clusters present, but in each
a similar front propagation of the infection appears.

\begin{figure}
  \includegraphics[width=\columnwidth]{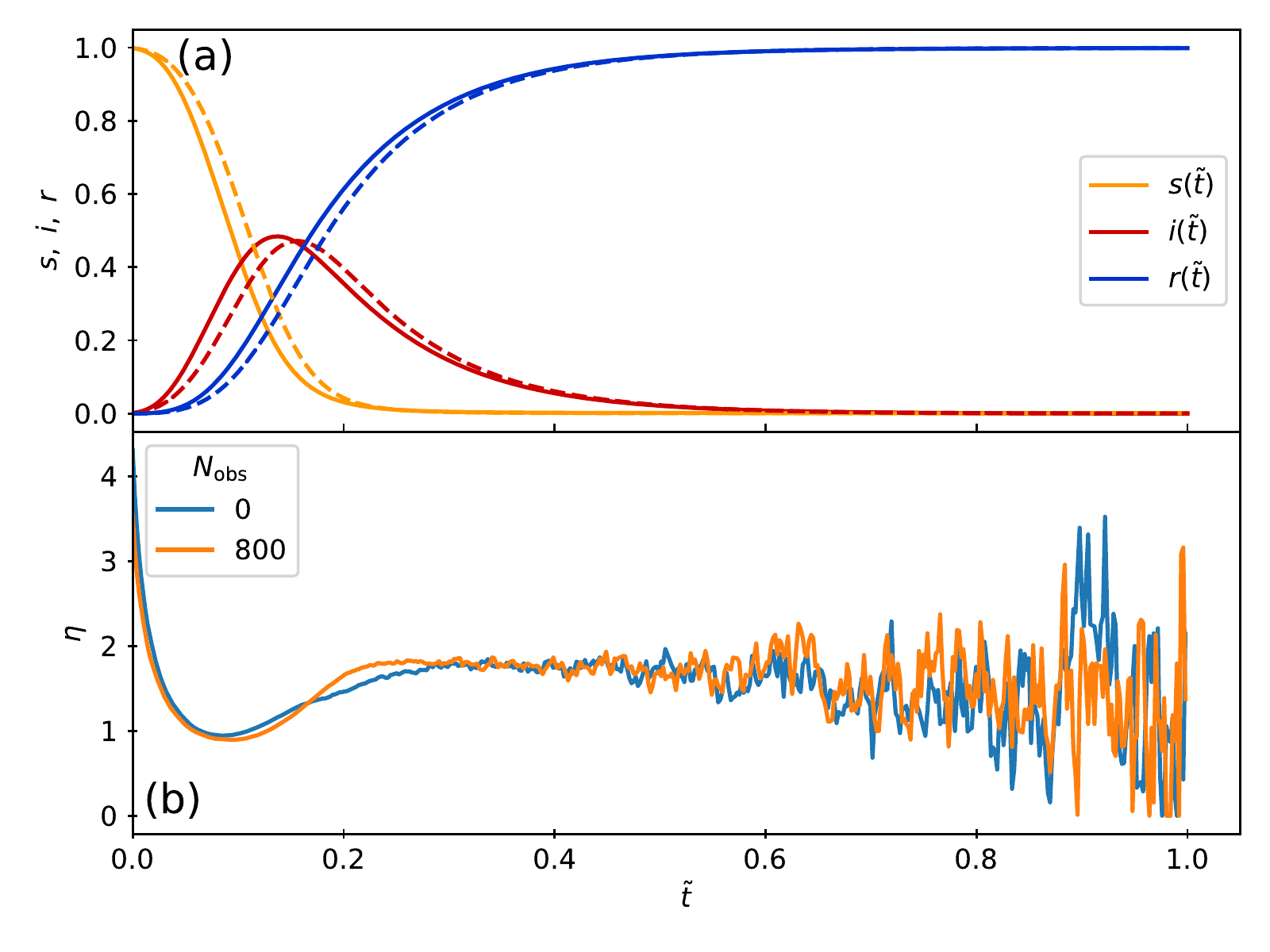}
  \caption{
{\bf Epidemic curves in the high transmissibility regime.}    
Fractions of susceptible $s$ (yellow), infected $i$ (red), and recovered $r$ (blue)
particles versus reduced time $\tilde{t}$ for the system in Fig.~\ref{fig:4} with
$\beta/\mu=5.0$.
Solid lines are for samples without quenched disorder and
dashed lines are for samples containing obstacles.
In this case all of the particles become infected and  $s_{\infty}=0$.
(b) The corresponding $\eta$, the average number of $S$ particles surrounding
an $I$ particle, versus $\tilde{t}$ in the samples without (blue) and with (orange)
obstacles.
There is a minimal difference in $\eta$ between the two cases.
  }
  \label{fig:5}
\end{figure}

\begin{figure}
  \includegraphics[width=\columnwidth]{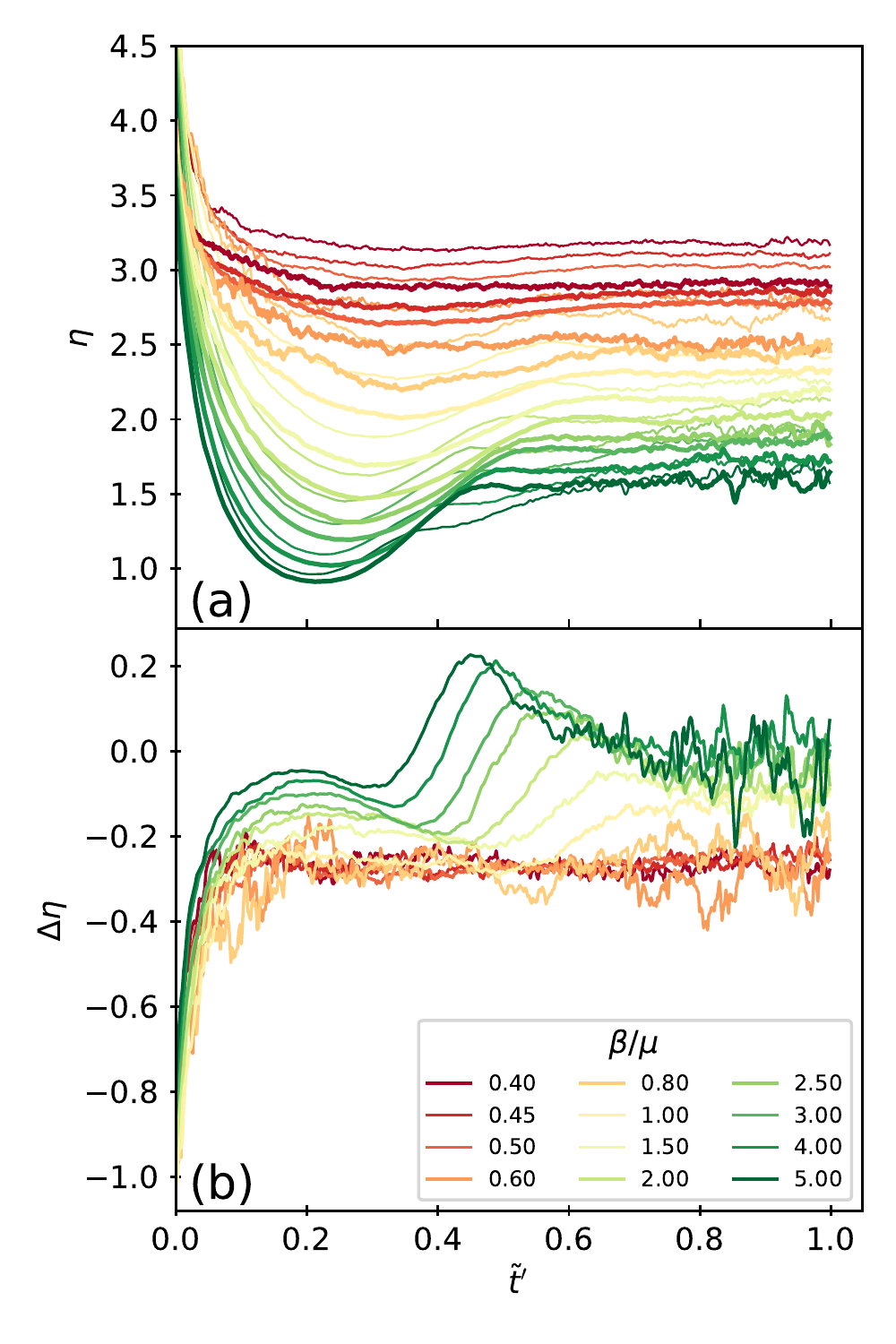}
  \caption{
    {\bf Measure of the ability of I to contact S and how it changes when quenched
    disorder is introduced.}    
(a) $\eta$ vs $\tilde{t}^\prime$ for varied $\beta/\mu$
in samples without obstacles (thin lines) 
and with obstacles (thick lines). When $\beta/\mu$ is large, a local minimum
in $\eta$ appears near $\tilde{t}^\prime=0.1$ due to the formation of a
propagating front.
(b)
The difference $\Delta \eta$ between the value of $\eta$ in the sample
with disorder and the value in the sample without disorder as a function of $\tilde{t}^\prime$.
For  $\beta/\mu \leq 1.5$, 
there is no front propagation and
the addition of quenched disorder always reduces the value of $\eta$.
For $\beta/\mu > 1.5$, a front appears, and once the front has passed,
$\Delta \eta$ drops below zero, indicating an enhancement of the
infection rate when quenched disorder is present.
  }
  \label{fig:7}
\end{figure}

\begin{figure}
  \includegraphics[width=0.5\textwidth]{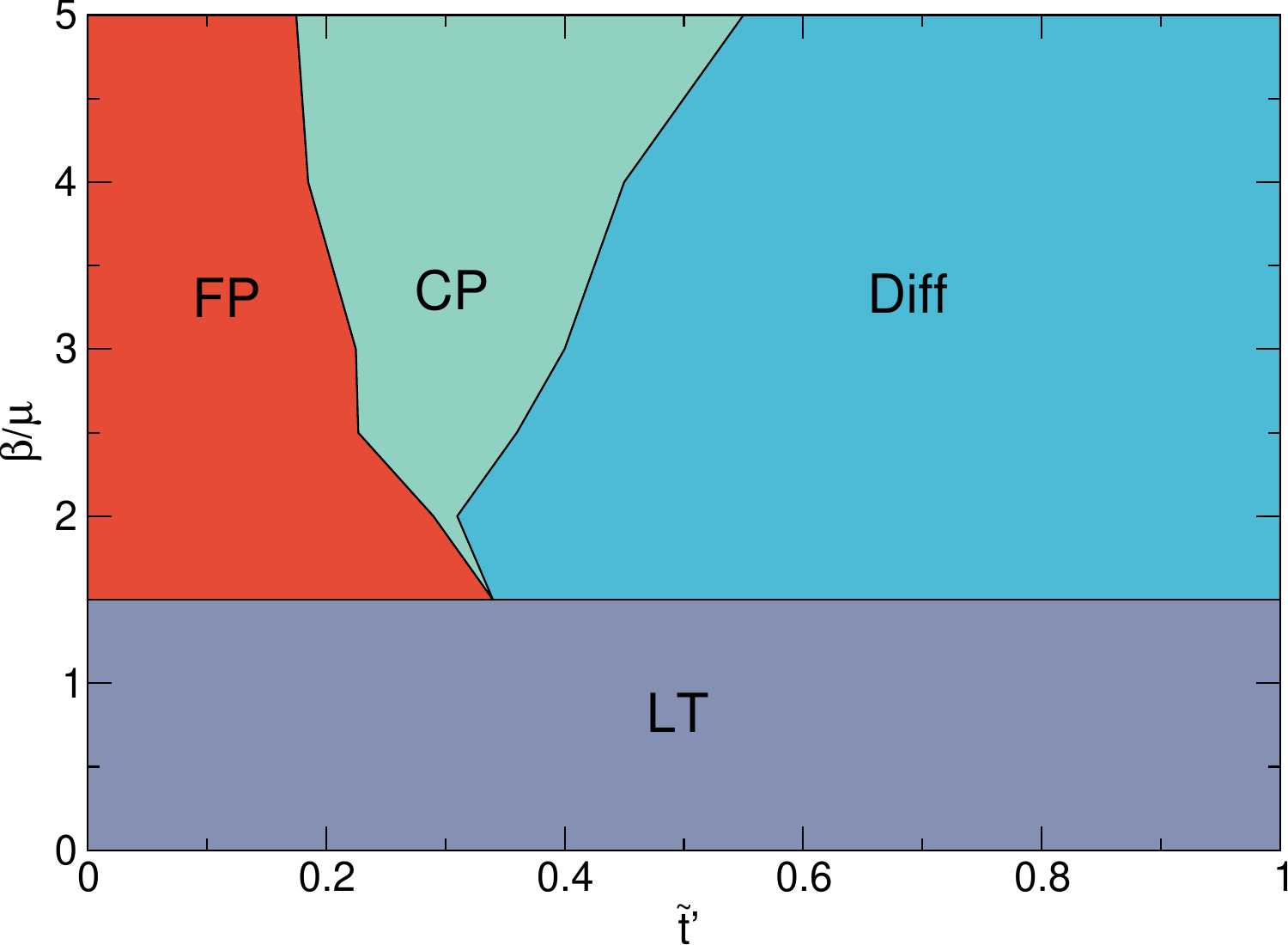}
  \caption{
    {\bf Phase diagram of the epidemic evolution in the low and high
    transmissibility regimes.}    
The phase diagram of the different regimes as a function of
transmissibility $\beta/\mu$ vs reduced time $\tilde{t}^\prime$ based on
the features of the curves in Fig.~\ref{fig:7} .
For $\beta/\mu \leq 1.5$, the system is in a low transmissibility (LT) regime
where $s_{\infty}>0$ and
the addition of obstacles can strongly impact the propagation of the
epidemic.
The front propagation phase at high transmissibility is
marked FP,
and in the secondary cluster phase (CP), the addition of
obstacles actually increases
the spread of the infection.
In the diffusive regime (Diff),  the obstacles do not affect 
the epidemic spread.
  }
  \label{fig:8}
\end{figure}

In Fig.~\ref{fig:5}(a) we plot $s(\tilde t)$, $i(\tilde t)$, and $r(\tilde t)$
for the
high transmissibility system with $\beta/\mu=5.0$ from Fig.~\ref{fig:4}.
Here, $s_{\infty}=0$ and all of the particles become infected regardless of
whether obstacles are present.
The peak value $i_{\rm max}$ is nearly the same for both cases.
An interesting effect appears in which
for $\tilde{t} < 0.175$, adding obstacles depresses $i$, but for
$\tilde{t} > 0.185$, adding obstacles increases $i$.
This is not merely due to a change in the duration of the epidemic since the curves
are plotted in reduced time; instead, it indicates a change in the spatial propagation
of the infection,
which we will address in Figs.~\ref{fig:7} and \ref{fig:8}.
The crossover in behavior
occurs after the initial large infection front has completely swept through
either the giant cluster or all of the smaller clusters for the samples with
quenched disorder.
In Fig.~\ref{fig:5}(b) we plot the corresponding $\eta$ versus $\tilde{t}$, which
is nearly unchanged by the inclusion of obstacles.
These results indicate that under high transmissibility,
the system is less sensitive to spatial disorder
and the behavior is
consistent with the mean field limit.
Note that epidemic curves and plots of $\eta(\tilde{t})$ for all other values of
$\beta/\mu$ can be viewed in the Supplemental Material.

\begin{figure}
  \includegraphics[width=0.9\columnwidth]{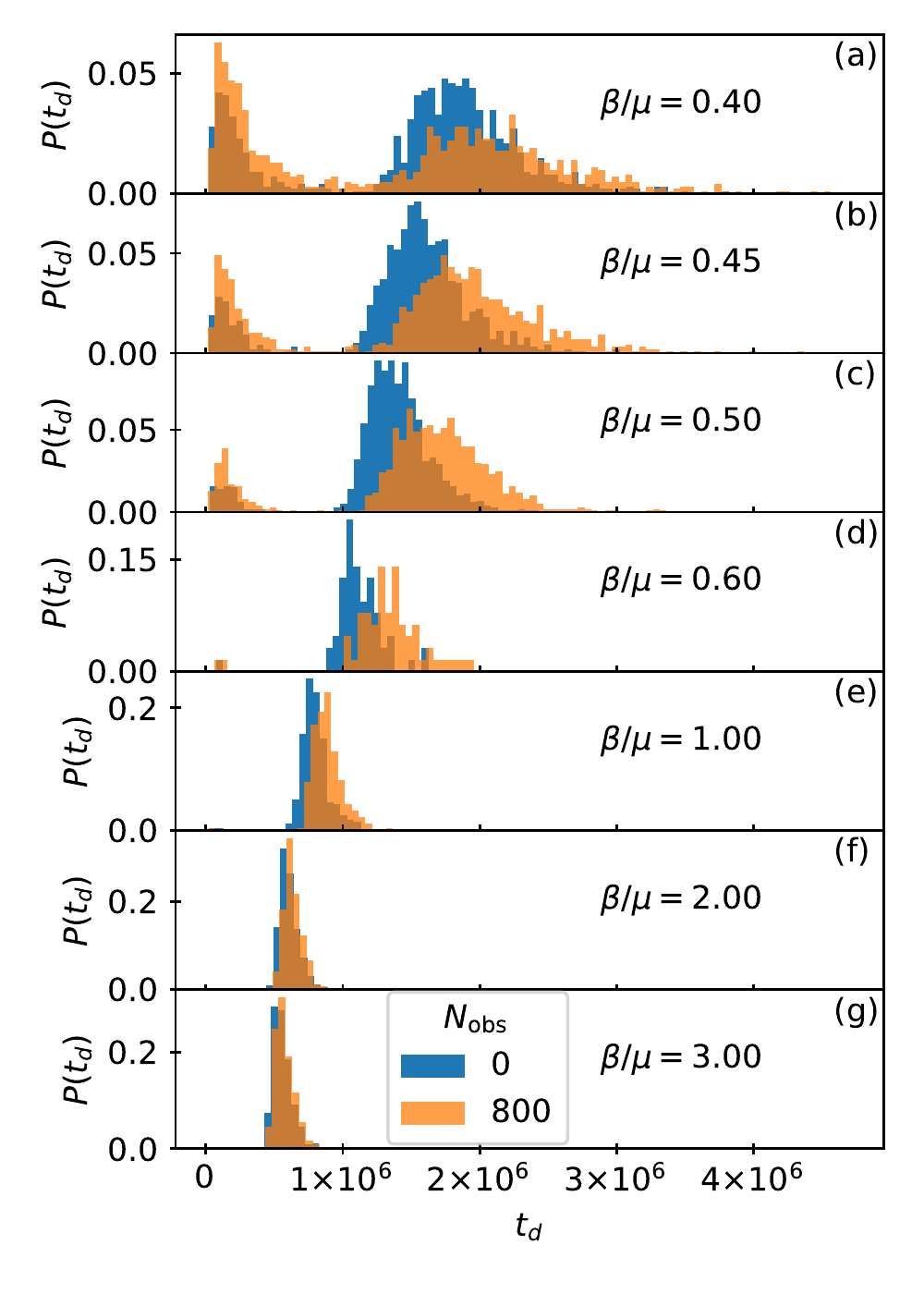}
  \caption{
{\bf Duration of epidemics with and without quenched disorder in the low
and high transmissibility regimes.}    
Distribution $P(t_d)$ of the duration $t_d$ of the epidemic in simulation
time steps
for 1000 realizations. Blue curves are for a system with no obstacles and
orange curves are for a system with obstacles.
The low transmissibility regime is
$\beta/\mu=$ (a) 0.4, (b) 0.45, (c) 0.5, (d) 0.6, and (e) 1.0, while
the high transmissibility regime is
$\beta/\mu=$ (f) 2.0 and (g) 3.0.
The distributions (a-e) in the low transmissibility regime are bimodal, and
the addition of quenched disorder increases the
number of failed outbreaks and increases the duration
of successful epidemics.
In the high transmissibility regime (f,g),
there are no failed outbreaks
and the effect of quenched disorder is strongly reduced.  
  }
  \label{fig:6}
\end{figure}

\noindent{\textbf{\textsf{Duration of Epidemic}}}\\
Our simulations reveal a strong stochasticity
of the behavior. Depending on the particular randomly chosen locations of the
index cases, the duration $t_d$ of the epidemic can vary widely. In particular,
for some realizations the outbreak fails to take hold and is extinguished without
affecting a significant fraction of the particles. To illustrate this, in Fig.~\ref{fig:6}
we plot the distribution $P(t_d)$ of the epidemics measured in simulation time
steps with and without obstacles
for 1000 realizations.
In Fig.~\ref{fig:6}(a-e),
we show the low transmissibility regime with $\beta/\mu=0.4, 0.45$, $0.5$, 0.6, and 1.0.
Here the distribution is bimodal and there is
a clear division between small $t_d$, where we find failed outbreaks that do
not infect a significant fraction of the particles, and larger $t_d$, where
successful epidemics occur that involve a substantial fraction of the particles.
This behavior is similar to what
has been observed in other studies \cite{Rock14}.
Addition of quenched disorder
in this regime increases the probability that the outbreak will fail, but
also increases the average duration of successful epidemics.
In contrast, for high transmissibility, as shown
in Fig.~\ref{fig:6}(f,g) at $\beta/\mu = 2.0$ and $3.0$,
$P(t_d)$ is unimodal since all outbreaks produce successful epidemics.
Additionally,
there is no longer a significant
difference in the distribution for systems with and without quenched disorder.

\noindent{\textbf{\textsf{Ability of I to Contact S}}}\\
We can also distinguish the two regimes of behavior using features in $\eta$
by comparing the value of $\eta$ in samples with and without quenched disorder.
In Fig.~\ref{fig:7}(a) we plot $\eta$ versus $\tilde{t}^\prime$ in samples with and without
obstacles.
The time scale $\tilde{t}^{\prime}$ reaches a value of
$\tilde{t}^{\prime}=1.0$ when the number of recovered
has increased to 95\% of its maximum value,
$r(\tilde{t}^\prime=1.0)=0.95r_{\infty}=0.95(1-s_{\infty})$. Use of this time scale allows us to exclude
the
stochastic late time behavior when the last few straggling infectives are recovering.
At $\tilde{t}^\prime = 0$, $\eta$ is always high since the initial seed $I$ particles
are surrounded only by susceptible particles. As
the epidemic spreads, the average number of $S$
particles
around $I$
particles decreases.
When $\beta/\mu \leq 1.5$,
the curves monotonically decrease to a saturation value between
$\eta=2.5$ to $\eta=3.25$, and samples containing obstacles
show lower values of $\eta$.
For $\beta/\mu > 1.5$, the epidemic spreads in the form of a front,
which is visible as
the appearance of a
local dip in $\eta$ centered near $\tilde{t}^\prime=0.2$.
As the front moves rapidly through the largest cluster, most of the infected
particles are surrounded by $I$ particles behind the expanding front,
while only $I$ particles at the edge of the front are adjacent to $S$ particles.
This depresses the value of $\eta$.
Once the front has passed through the cluster, the mobility of the particles
bring more $S$ from the gas phase into contact with the remaining $I$, and
$\eta$ recovers somewhat before saturating to a low value between
$\eta=1.5$ and 2.0.

In Fig.~\ref{fig:7}(b) we plot the difference
$\Delta \eta=\eta_{\rm obs=800}-\eta_{\rm obs=0}$
between $\eta$ for the samples with and without quenched disorder
from Fig.~\ref{fig:7}(a). 
When $\beta/\mu \leq 1.5$,
$\Delta \eta$
reaches a constant value and is always
negative,
$\Delta \eta \approx -0.25$,
indicating that the quenched disorder is always reducing
the effectiveness of the spread of the epidemic. 
Once the system enters the front propagation regime for
$\beta/\mu > 1.5$, $\Delta \eta$ becomes nonmonotonic and
shows local peaks and dips.
For $\tilde{t}^\prime<0.4$, there is a dip
when the front is passing through the largest clusters.
In this regime, $\Delta \eta$ is negative,
indicating that the quenched disorder slows the front to some extent.
For times above the minimum of the dip,
$\Delta \eta$ increases and becomes
positive, indicating that the addition of quenched disorder
is actually increasing the effectiveness of the epidemic spread.
This can also be seen in Fig.~\ref{fig:5}(a), where $i$ is reduced in the presence
of quenched disorder
for $\tilde{t} < 0.175$ but is slightly increased
for $\tilde{t} > 0.175$, indicating that the disorder can accelerate the 
infection at later times.
This enhancement of the epidemic arises after the largest cluster has become
fully infected and some of the infected particles break away from the
cluster and enter the gas phase.
Within the gas phase, the quenched disorder induces the formation of smaller
localized clusters, as shown
in Fig.~\ref{fig:1}.
These smaller clusters, once contacted by an infective, undergo
the same rapid front propagation as the initial infection wave.
When quenched disorder is not present, there are no smaller clusters and
the infection must propagate through the gas phase and infect the remaining
$S$ particles one by one, an inefficient process.

\noindent{\textbf{\textsf{Epidemic Phase Diagram}}}\\
Based on the features in Fig.~\ref{fig:7}, we can construct a phase diagram of
the behavior of the system as a function of
$\beta/\mu$ versus $\tilde{t}^\prime$,
as shown in Fig.~\ref{fig:8}.
When $\beta/\mu > 1.5$,
$s_{\infty}=0$ and the entire system becomes infected, while
the initial invasion of the infection
occurs by front propagation.
For $\beta/\mu \leq 1.5$, the low transmissibility regime marked LT,
the infection spreads much more homogeneously,
as illustrated
in Fig.~\ref{fig:2}, and $s_{\infty}>0$ so that
not all of the particles have been
infected by the end of the epidemic.
Within this regime, the addition of quenched disorder
always reduces $i_{\rm max}$ and increases $s_{\infty}$.
In the high transmissibility regime with
$\beta/\mu > 1.5$,
at small $\tilde{t}^\prime$ the infection propagates
as a front through the largest cluster, defined as the front propagation regime FP.
Here the addition of quenched disorder can slow the front propagation but does
not stop it.
After the front has crossed the entire largest cluster, we find the CP regime
in which
the secondary clusters start to show front propagation.
Here the quenched disorder can increase the effectiveness of the spread of
the infection by increasing the number of secondary clusters
that are present.
At larger values of $\tilde{t}^\prime$, all of the clusters have been infected and
the epidemic is making its way through the gas phase.
In this regime, which we call Diff for diffusive,
there is little difference 
between the systems with and without quenched disorder.
The locations of the phase boundaries should
depend on the amount of quenched disorder and the activity level of the particles. 

\section{Discussion}

Our model suggests that active matter can be used to capture a variety 
of epidemic behaviors.
There are a number of active matter systems, such as active colloids, in 
which the activity of the particles can be controlled on an individual basis
using optical rastering methods.
Experiments of this type have been developed in order
to use active colloids to mimic group formation,
to introduce an effective visual perception mobility,
and to produce other kinds of
collective behaviors such as quorum sensing \cite{Lavergne19,Bauerle20}.
In order to implement an S-I-R model,
individual active colloids could be
traced and tagged according to their infective state,
and when they interact with other colloids,
there can be a probability that the infection will pass to a susceptible colloid.
This can be done
in a motility induced phase separated regime or in a diffusive regime
for varied $\beta/\mu$.
The experiments could then be repeated multiple times to obtain
the average behavior.
Within a given sample,
certain colloids could remain inactive and be counted as passive or obstacle particles,
or actual obstacles could be put in place on the substrate.
Additionally, a wealth of rules could be introduced, such as
the inclusion of hyperactive particles
that could serve as superspreaders, as well as
possible mitigation effects.
This could potentially position
active matter as a
table-top experimental system for modeling
epidemics.
Our work indicates that  active matter
can be used as a simulation tool to study epidemics in a system that can
be tuned readily between states that are sensitive to spatial disorder and
states that are insensitive to disorder.

In conclusion, 
we have shown how an active matter system of self-propelled particles 
can be used to model spatial heterogeneity in
an S-I-R epidemic spreading model.
The active particles naturally form spatial clusters
in the motility induced phase separated regime. 
For low transmissibility,
the epidemic spread is percolative and the system is sensitive to the
addition of quenched disorder, which both
increases the probability of a failed outbreak and
increases the average duration of successful epidemics.
In this regime,
the mixing hypothesis of classical S-I-R models
breaks down.
For high transmissibility, all of the particles are eventually infected
and the epidemic spreads
in well defined fronts.
In this case, the addition of quenched disorder can
slow the spread of the epidemic at early times by slowing the propagation of the
initial front. At later times, however,
since the quenched disorder introduces a larger number of small clusters in the
gas phase,
the epidemic can spread more efficiently compared to a system without
quenched disorder.
Our results indicate
that spatial disorder can impact epidemic spreading in 
both the high and low transmissibility regimes.
Our system could be realized experimentally using
light activated colloidal particles with
specified feedback rules to mimic the S-I-R model,
and our results suggest that active matter
systems could provide a new way
to create table-top epidemic experiments.

\smallskip

\noindent{\textbf{\textsf{Acknowledgements}}}\\
This work was supported by the US Department of Energy through
the Los Alamos National Laboratory.  Los Alamos National Laboratory is
operated by Triad National Security, LLC, for the National Nuclear Security
Administration of the U. S. Department of Energy (Contract No. 892333218NCA000001).
NH benefited from resources provided by the Center for Nonlinear Studies (CNLS).
PF and AL were supported by a grant of the Romanian Ministry of Education
and Research, CNCS - UEFISCDI, project number
PN-III-P4-ID-PCE-2020-1301, within PNCDI III.

\bibliographystyle{naturemag}

\bibliography{mybib}

\vfill\eject
\setcounter{figure}{0}

{\bf Supplemental material for 
``Using Active Matter to Introduce Spatial Heterogeneity to the Susceptible-Infected-Recovered Model of Epidemic Spreading'' 
}

Here we present the full epidemic curves of susceptible $s(\tilde{t})$,
infected $i(\tilde{t})$, and recovered
$r(\tilde{t})$ along with the
corresponding $\eta(\tilde{t})$, the average number of $S$ particles surrounding
an $I$ particle, for additional values of $\beta/\mu$ that were
not included in the main text.
Each curve is averaged over 1000 realizations.
At $\tilde{t} = 1.0$, the epidemic is over and $i=0$.
\vspace{-0.2in}

\begin{figure}[H]
  \includegraphics[width=0.45\textwidth]{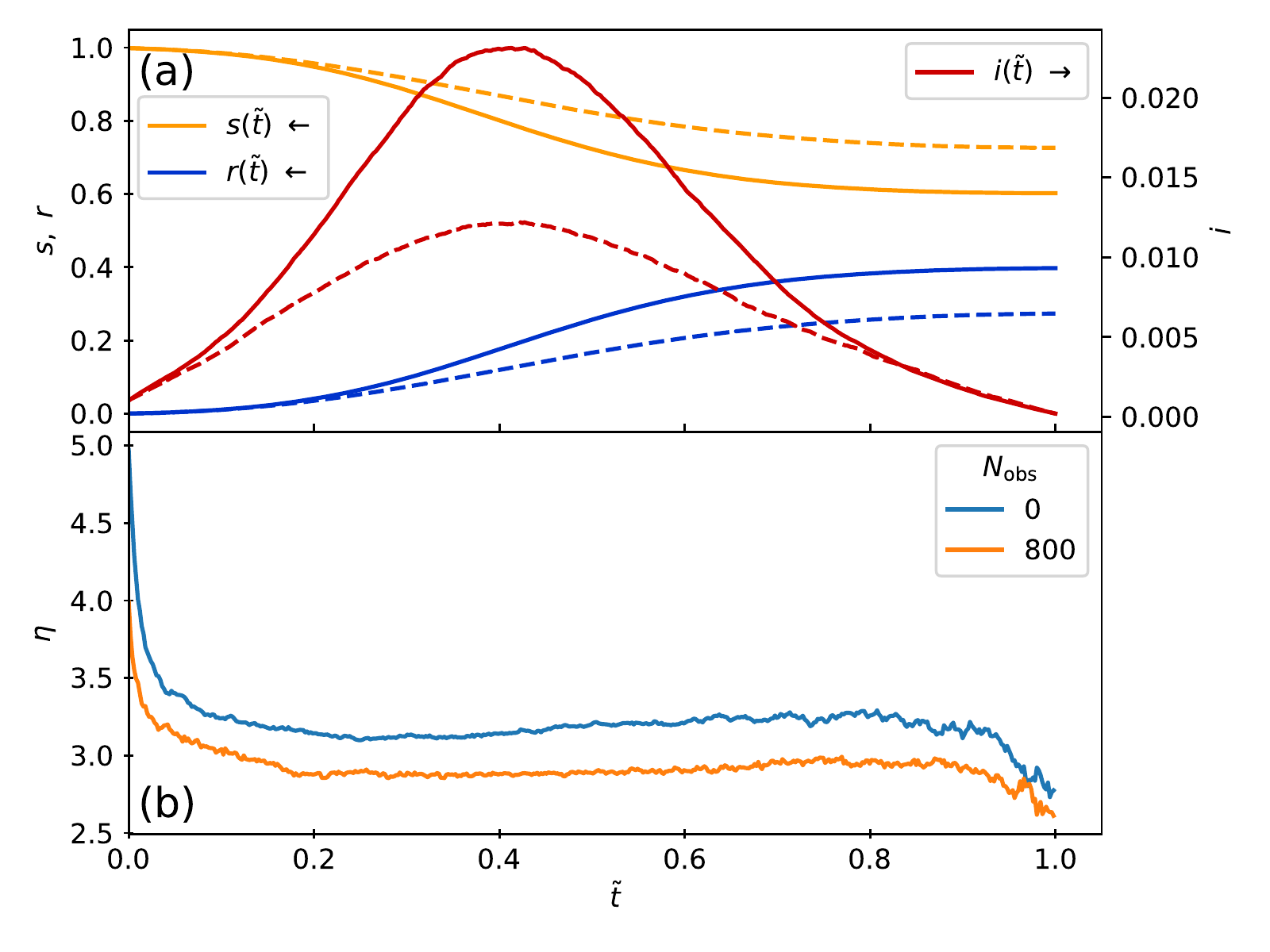}
  \vspace{-0.3in}
  \caption{
{\bf Epidemic curves in the low transmissibility regime.}
(a) $s(\tilde{t})$ (yellow), $i(\tilde{t})$ (red), and $r(\tilde{t})$ (blue)
for a system with $\beta/\mu=0.4$.
Solid lines: no quenched disorder; dashed lines: samples containing obstacles.
(b) The corresponding $\eta(\tilde{t})$
without (blue) and with (orange) obstacles.
}
  \label{fig:1s}
\end{figure}
\vspace{-0.2in}

\begin{figure}[H]
  \includegraphics[width=0.45\textwidth]{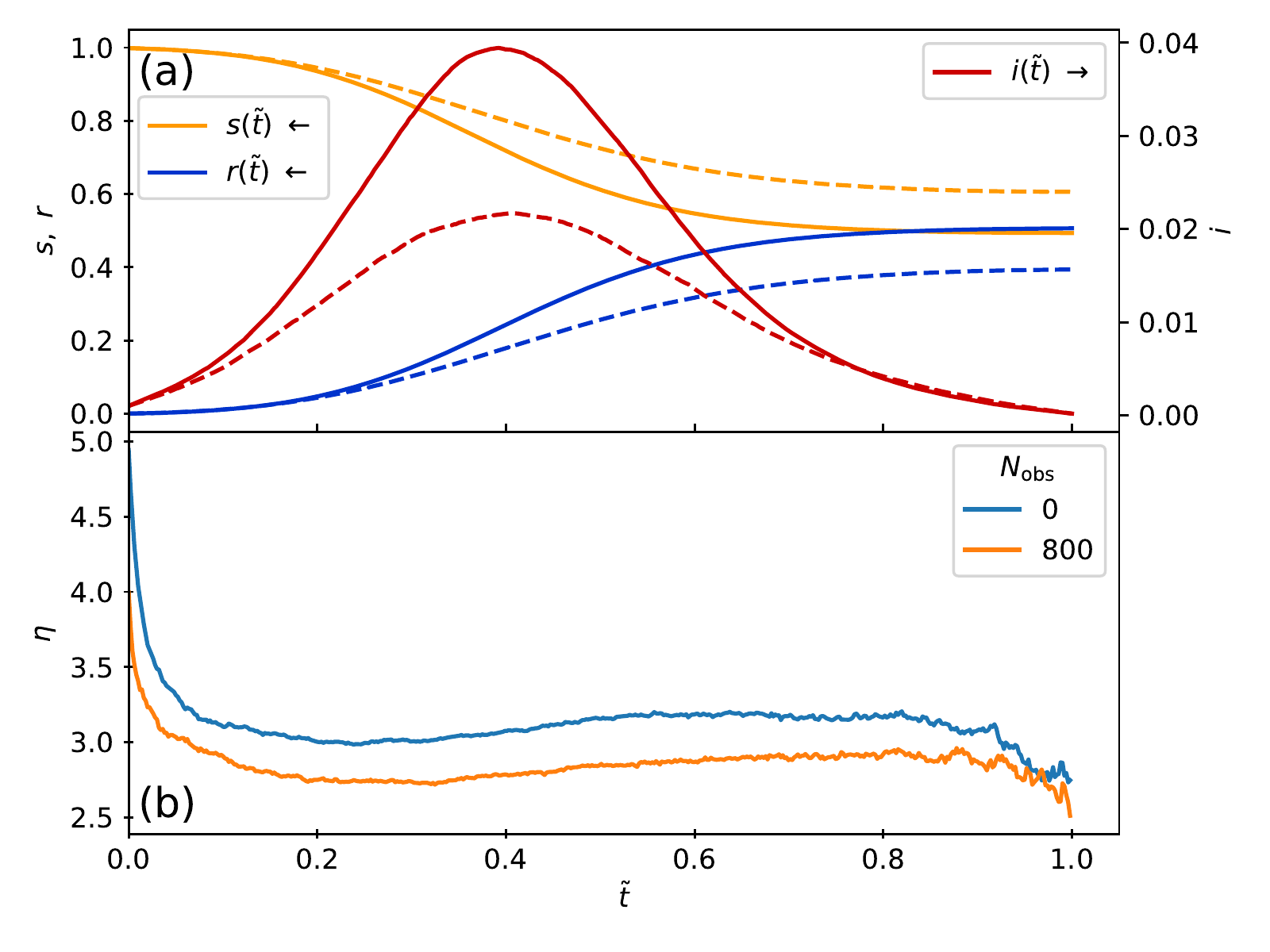}
  \vspace{-0.3in}
  \caption{
{\bf Epidemic curves in the low transmissibility regime.}
(a) $s(\tilde{t})$ (yellow), $i(\tilde{t})$ (red), and $r(\tilde{t})$ (blue)
for a system with $\beta/\mu=0.45$.
Solid lines: no quenched disorder; dashed
lines: samples containing obstacles.
(b) The corresponding $\eta(\tilde{t})$
without (blue) and with (orange) obstacles.
}
  \label{fig:2s}
\end{figure}

\begin{figure}[H]
  \includegraphics[width=0.45\textwidth]{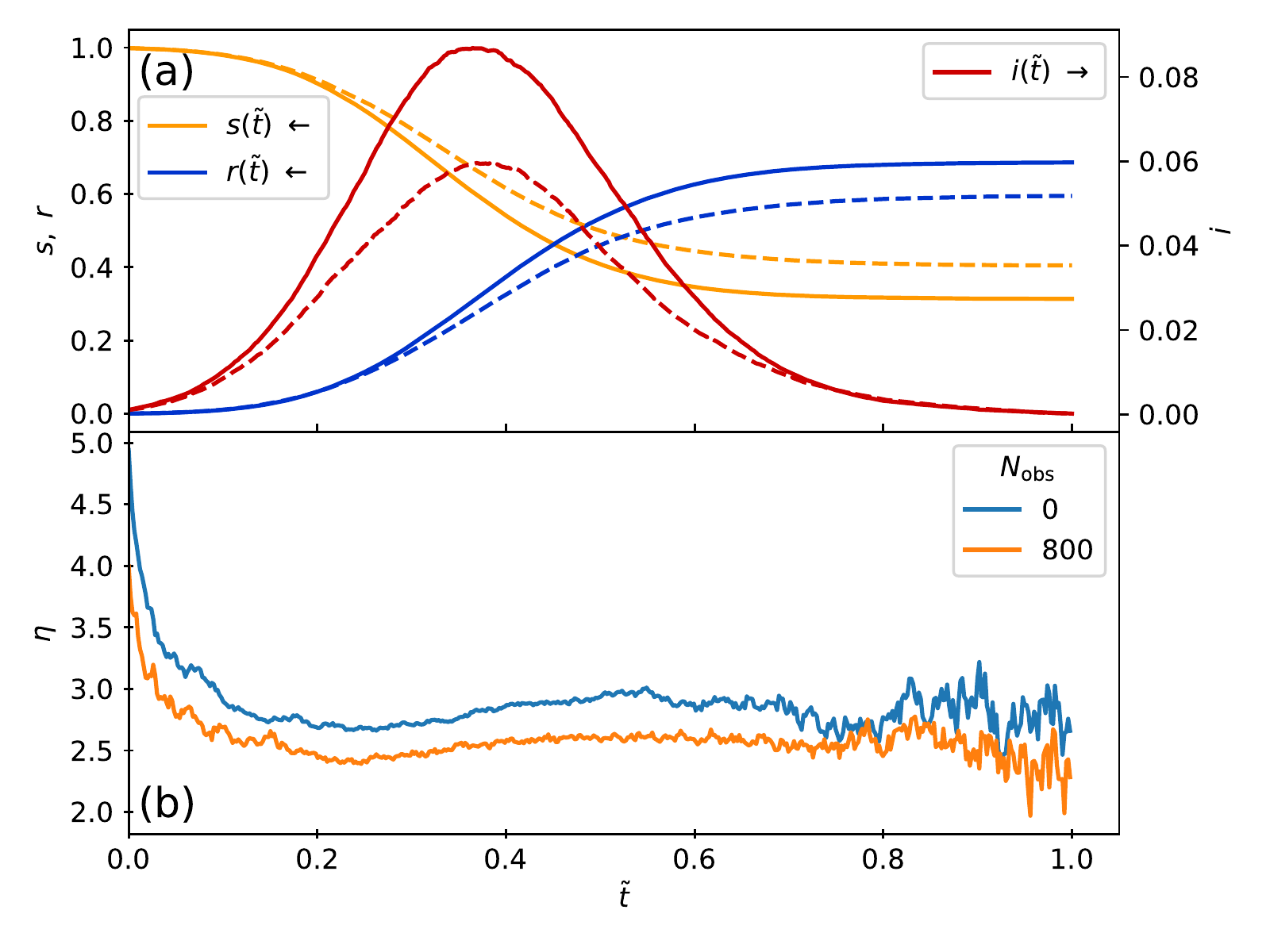}
  \vspace{-0.3in}
  \caption{
{\bf Epidemic curves in the low transmissibility regime.}
(a) $s(\tilde{t})$ (yellow), $i(\tilde{t})$ (red), and $r(\tilde{t})$ (blue)
for a system with $\beta/\mu=0.6$.
Solid lines: no quenched disorder; dashed
lines: samples containing obstacles.
(b) The corresponding $\eta(\tilde{t})$
without (blue) and with (orange) obstacles.
}
  \label{fig:3s}
\end{figure}
\vspace{-0.2in}

\begin{figure}[H]
  \includegraphics[width=0.45\textwidth]{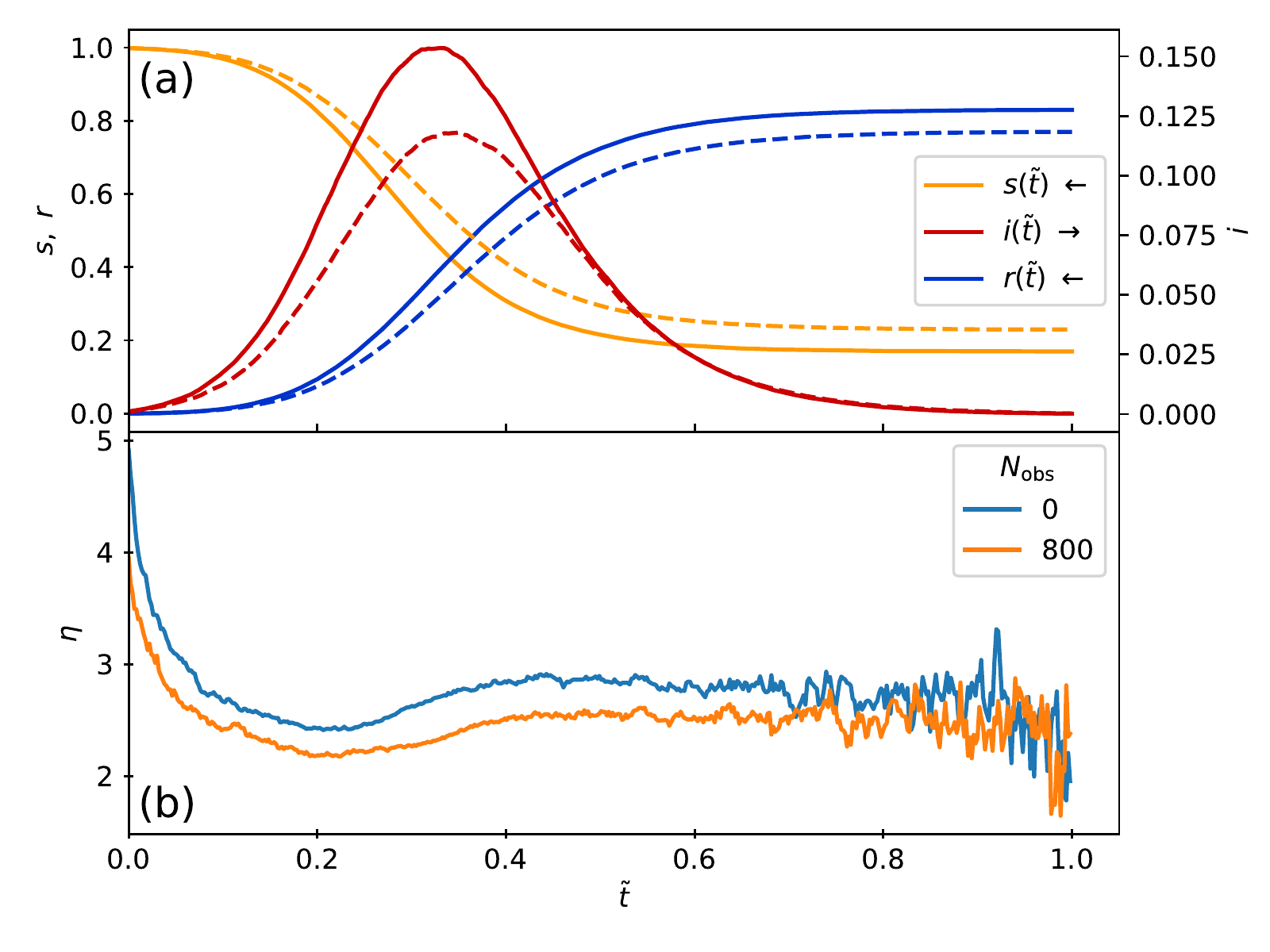}
  \vspace{-0.3in}
  \caption{
{\bf Epidemic curves in the low transmissibility regime.}
(a) $s(\tilde{t})$ (yellow), $i(\tilde{t})$ (red), and $r(\tilde{t})$ (blue)
for a system with $\beta/\mu=0.8$.
Solid lines: no quenched disorder; dashed
lines: samples containing obstacles.
(b) The corresponding $\eta(\tilde{t})$
without (blue) and with (orange) obstacles.
}
  \label{fig:4s}
\end{figure}
\vspace{-0.2in}

\begin{figure}[H]
  \includegraphics[width=0.45\textwidth]{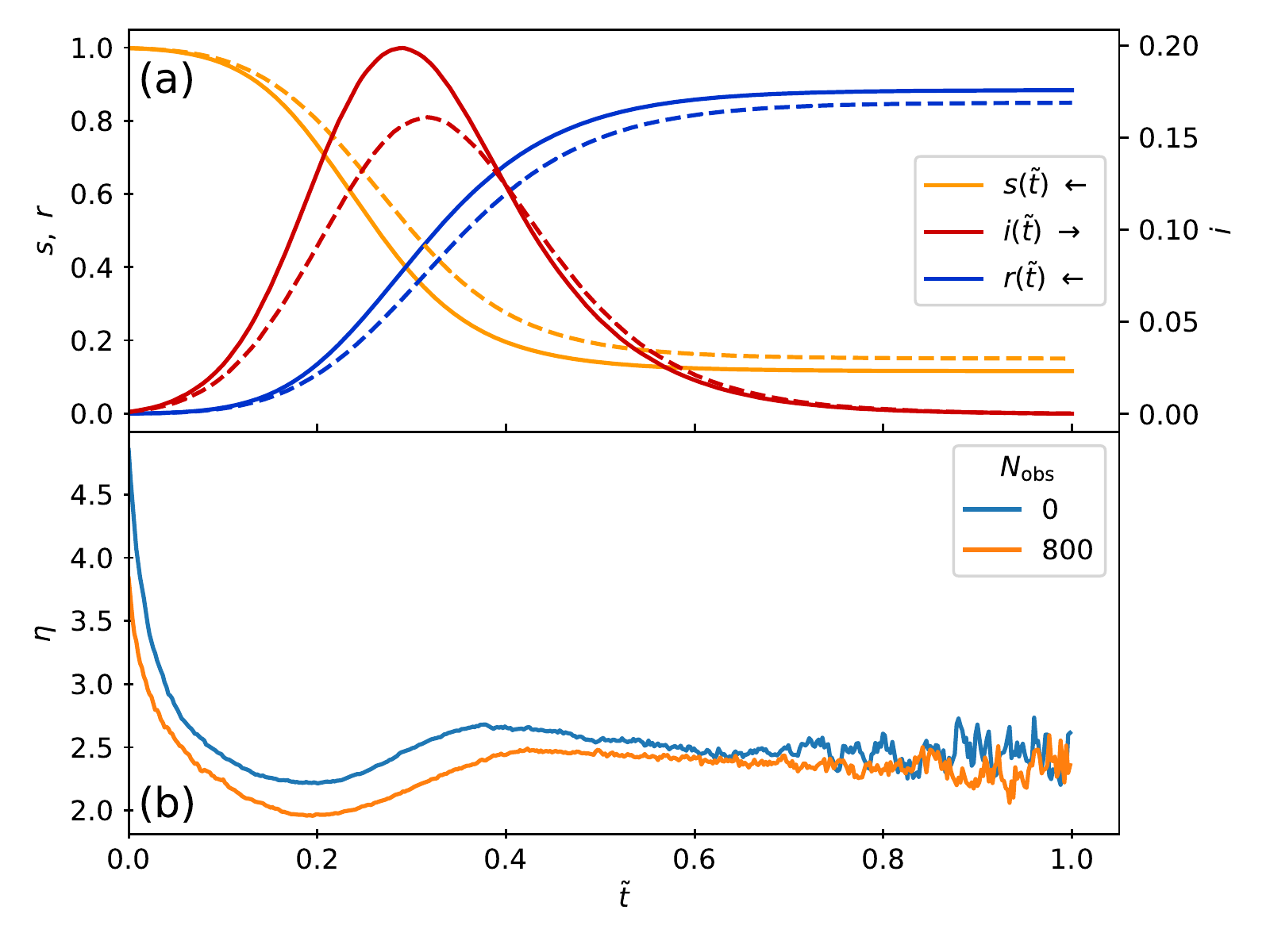}
  \vspace{-0.3in}
  \caption{
{\bf Epidemic curves in the low transmissibility regime.}
(a) $s(\tilde{t})$ (yellow), $i(\tilde{t})$ (red), and $r(\tilde{t})$ (blue)
for a system with $\beta/\mu=1.0$.
Solid lines: no quenched disorder; dashed
lines: samples containing obstacles.
(b) The corresponding $\eta(\tilde{t})$
without (blue) and with (orange) obstacles.
}
  \label{fig:5s}
\end{figure}
\vspace{-0.2in}

\begin{figure}[H]
  \includegraphics[width=0.45\textwidth]{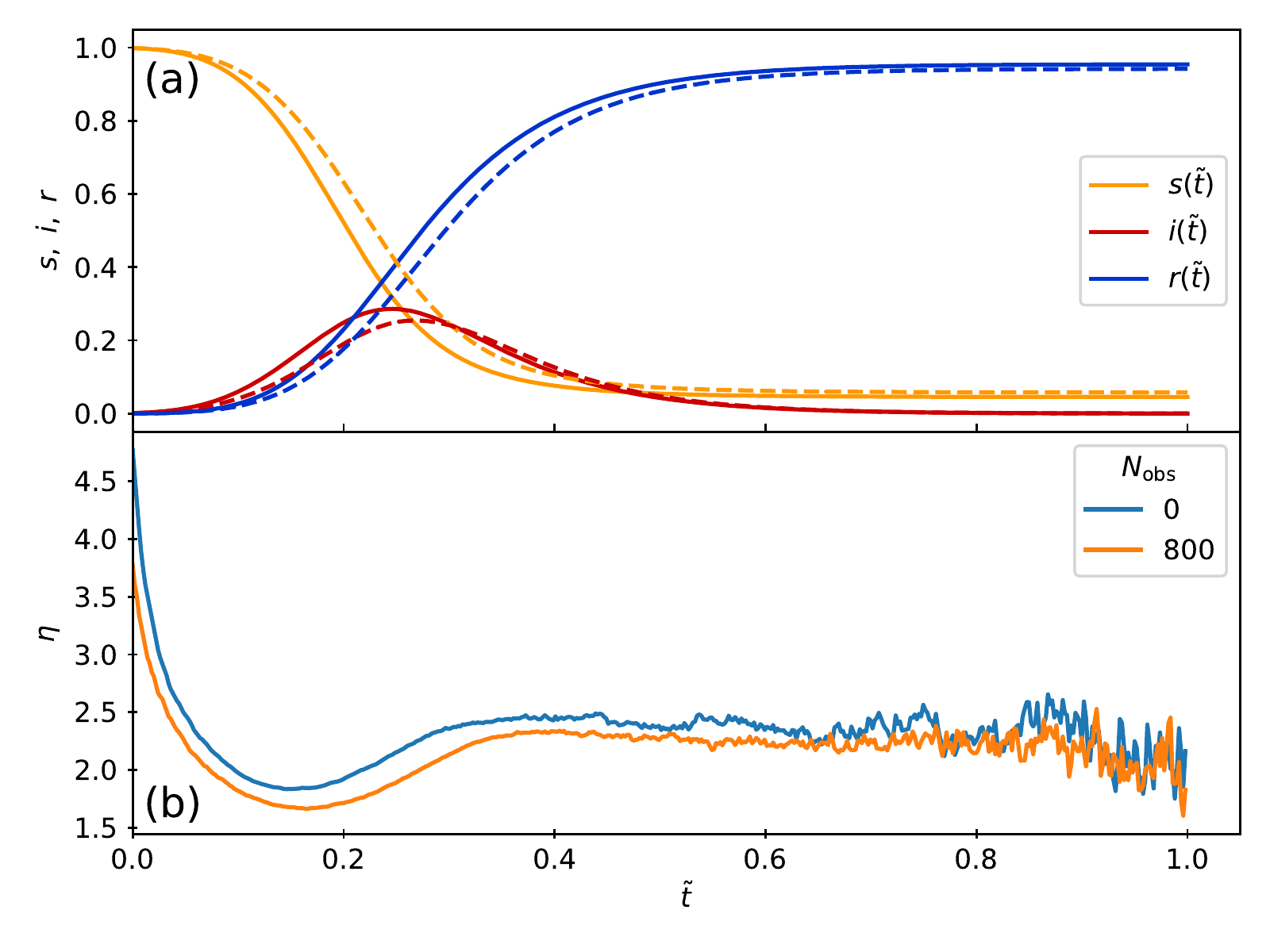}
  \vspace{-0.3in}
  \caption{
{\bf Epidemic curves in the transitional transmissibility regime.}
(a) $s(\tilde{t})$ (yellow), $i(\tilde{t})$ (red), and $r(\tilde{t})$ (blue)
for a system with $\beta/\mu=1.5$.
Solid lines: no quenched disorder; dashed
lines: samples containing obstacles.
(b) The corresponding $\eta(\tilde{t})$
without (blue) and with (orange) obstacles.
}
  \label{fig:6s}
\end{figure}
\vspace{-0.3in}

\begin{figure}[H]
  \includegraphics[width=0.45\textwidth]{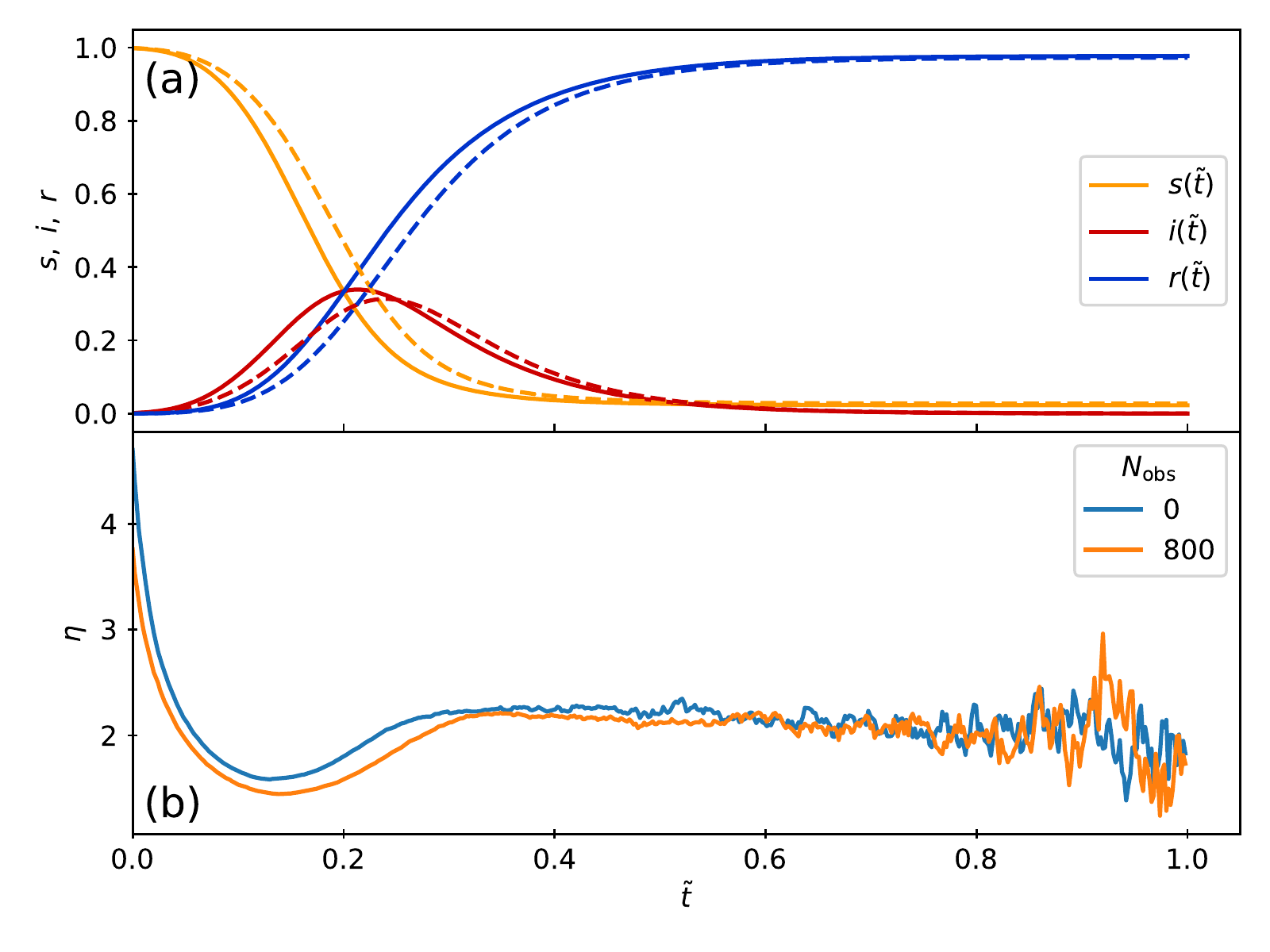}
  \vspace{-0.3in}
  \caption{
{\bf Epidemic curves in the high transmissibility regime.}
(a) $s(\tilde{t})$ (yellow), $i(\tilde{t})$ (red), and $r(\tilde{t})$ (blue)
for a system with $\beta/\mu=2.0$.
Solid lines: no quenched disorder; dashed
lines: samples containing obstacles.
(b) The corresponding $\eta(\tilde{t})$
without (blue) and with (orange) obstacles.
}
  \label{fig:7s}
\end{figure}
\vspace{0.5in}

\begin{figure}[H]
  \includegraphics[width=0.45\textwidth]{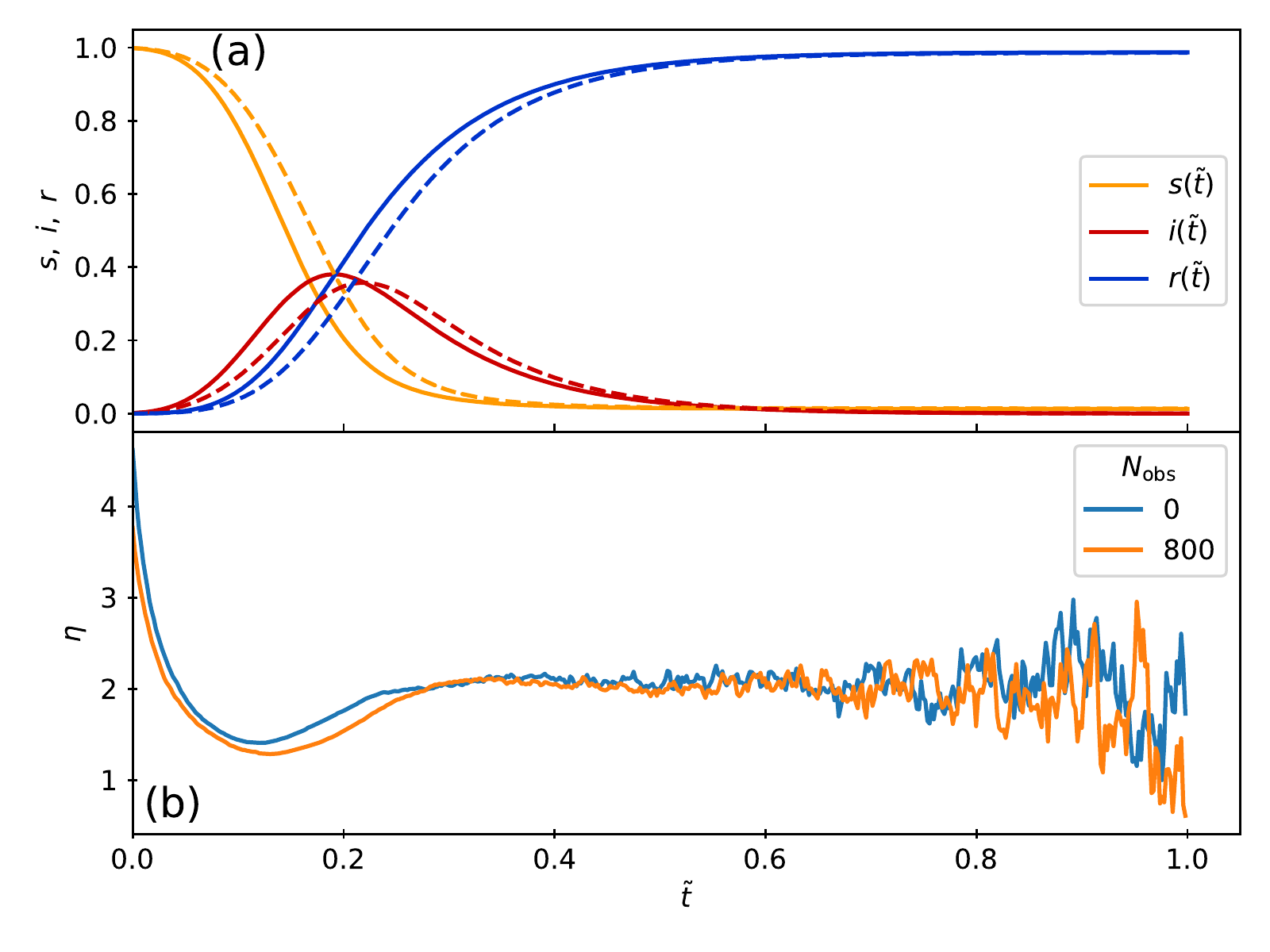}
  \vspace{-0.3in}
  \caption{
{\bf Epidemic curves in the high transmissibility regime.}
(a) $s(\tilde{t})$ (yellow), $i(\tilde{t})$ (red), and $r(\tilde{t})$ (blue)
for a system with $\beta/\mu=2.5$.
Solid lines: no quenched disorder; dashed
lines: samples containing obstacles.
(b) The corresponding $\eta(\tilde{t})$
without (blue) and with (orange) obstacles.
}
  \label{fig:8s}
\end{figure}
\vspace{-0.3in}

\begin{figure}[H]
  \includegraphics[width=0.45\textwidth]{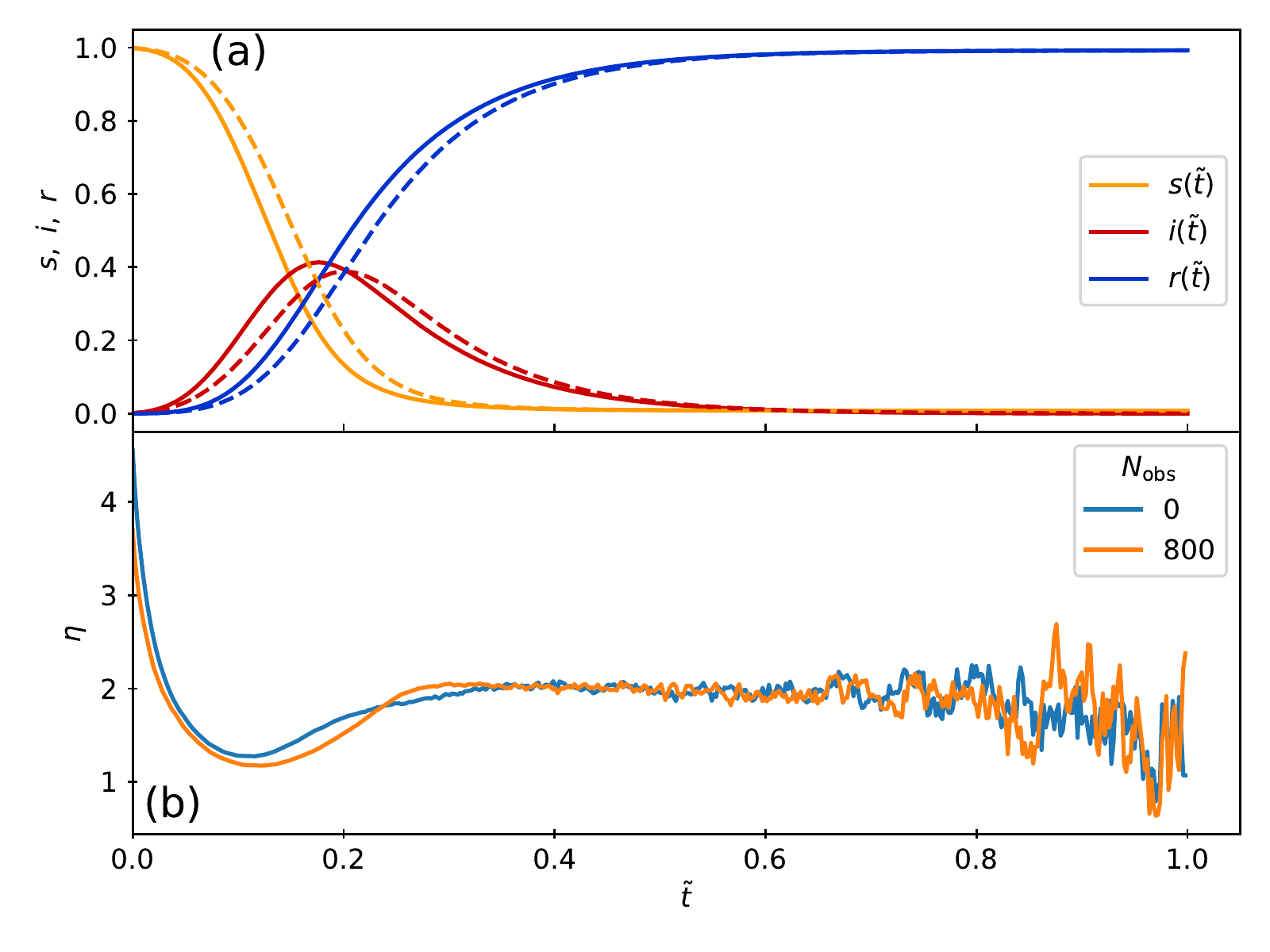}
  \vspace{-0.3in}
  \caption{
{\bf Epidemic curves in the high transmissibility regime.}
(a) $s(\tilde{t})$ (yellow), $i(\tilde{t})$ (red), and $r(\tilde{t})$ (blue)
for a system with $\beta/\mu=3.0$.
Solid lines: no quenched disorder; dashed
lines: samples containing obstacles.
(b) The corresponding $\eta(\tilde{t})$
without (blue) and with (orange) obstacles.
}
  \label{fig:9s}
\end{figure}
\vspace{-0.3in}

\begin{figure}[H]
  \includegraphics[width=0.45\textwidth]{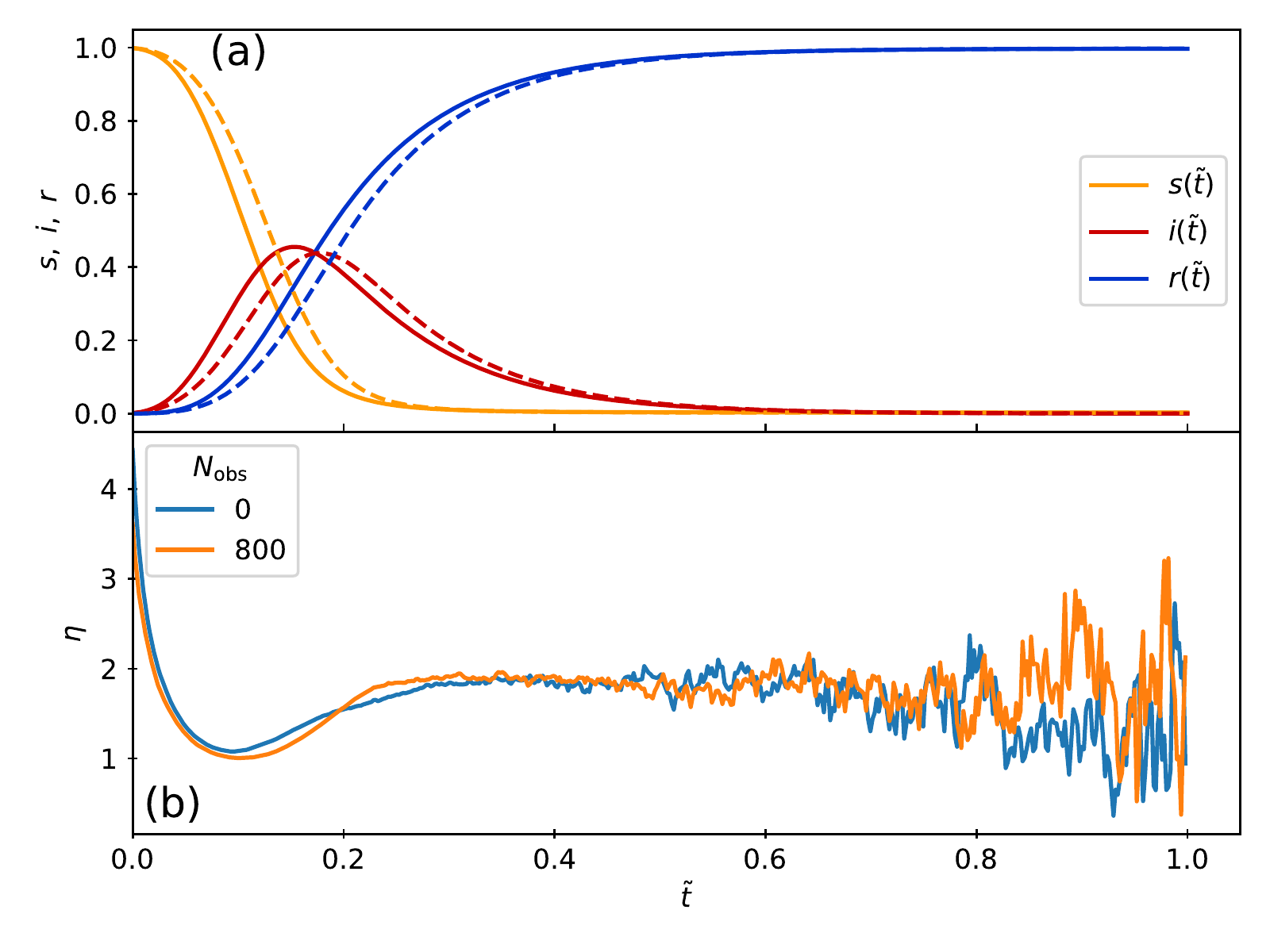}
  \vspace{-0.3in}
  \caption{
{\bf Epidemic curves in the high transmissibility regime.}
(a) $s(\tilde{t})$ (yellow), $i(\tilde{t})$ (red), and $r(\tilde{t})$ (blue)
for a system with $\beta/\mu=4.0$.
Solid lines: no quenched disorder; dashed
lines: samples containing obstacles.
(b) The corresponding $\eta(\tilde{t})$
without (blue) and with (orange) obstacles.
}
  \label{fig:10s}
\end{figure}

\end{document}